# Generative AI and Machine Learning Collaboration for Container Dwell Time Prediction via Data Standardization

Minseop Kim[a], Takhyeong Kim[b], Taekhyun Park[b], Hanbyeol Park[a], Hyerim Bae[b]

[a]Major in Industrial Data Science & Engineering. Department of Industrial Engineering, Pusan National University, Republic of Korea

[b]Graduate School of Data Science, Pusan National University, Republic of Korea

**Abstract**

Import container dwell time (ICDT) prediction is a key task for improving productivity in container terminals, as accurate predictions enable the reduction of container rehandling operations by yard cranes. Achieving this objective requires accurately predicting the dwell time of individual containers. However, the primary determinants of dwell time are recorded as unstructured text, which limits their effective use in machine learning models. This study addresses this limitation by proposing a collaborative framework that integrates generative artificial intelligence (Gen AI) with machine learning. The proposed framework employs Gen AI to standardize unstructured information into standard international codes, with dynamic re-prediction triggered by electronic data interchange state updates, enabling the machine learning model to predict ICDT accurately. Extensive experiments conducted on real container terminal data demonstrate that the proposed methodology achieves a 13.88% improvement in mean absolute error compared to conventional models that do not utilize standardized information. Furthermore, applying the improved predictions to container stacking strategies achieves up to 13.07% reduction in the number of relocations, thereby empirically validating the potential of Gen AI to enhance productivity in container terminal operations. Overall, this study provides both technical and methodological insights into the adoption of Gen AI in port logistics and its effectiveness.

## 1. Introduction

Container shipping accounts for more than 50% of global maritime trade, and container ports function as critical hubs linking maritime and inland transportation (Xie et al. 2025). Major ports worldwide handle hundreds of millions of containers annually, playing an essential role in the smooth functioning of global supply chains (Xie et al. 2017). Accordingly, container terminals responsible for unloading and storing containers have developed and implemented various methodologies to improve yard management efficiency (Filom et al. 2022; Kourounioti et al. 2016; K. Park et al. 2026). Among these methodologies, accurately identifying container dwell time (CDT) is widely recognized as an important process for enhancing container handling efficiency (Chhetri et al. 2024).

Containers handled at ports are generally classified into three types: import, export, and transshipment (Irannezhad et al. 2019). For export and transshipment cargo, CDT can be calculated based on scheduled vessel arrival times (Gaete et al. 2017; Rodriguez-Molins et al. 2012). Accordingly, operators establish stacking strategies in their daily operational decisions, such as placing containers scheduled for loading in marshalling yards close to berthing locations or stacking them in advance at upper tiers (Kourounioti et al. 2016). These practices are regarded as core processes for efficient container terminal operations, as they can reduce vessel handling time and decrease the number of relocations performed by yard cranes (De Armas Jacomino et al. 2021; Maldonado et al. 2019).

In contrast, for import containers unloaded from vessels, the arrival time of the pickup truck transporting the container is not known in advance (Ku and Arthanari 2016; Saurí and Martín 2011), making import container dwell time (ICDT) difficult to identify. This uncertainty poses a major constraint on the effective application of various yard operation strategies. Consequently, many studies have treated ICDT as a prediction problem, primarily employing machine learning (ML) and artificial intelligence (AI) techniques (Chhetri et al. 2024; Gaete et al. 2017; Kourounioti et al. 2016; Saini and Lerher 2024). However, existing studies suffer from a key limitation in that they fail to effectively utilize cargo information (CI) and owner information (OI), which are key determinants of ICDT (Hassan and Gurning 2020). These variables are recorded in unstructured textual form and cannot be directly used in ML models. Instead, relevant information must be manually interpreted and categorized according to standard classification systems by experts. Such manual standardization processes are time-consuming and costly, rendering information extraction highly impractical in operational settings (Xie et al. 2025). Nevertheless, transforming high-dimensional, unstructured port data into meaningful features has been consistently emphasized as a prerequisite for the effective application of AI in port logistics (Chung 2021; Filom et al. 2022).

To address these challenges, generative AI (Gen AI) has recently emerged as a promising technology. Owing to its advanced language understanding capabilities, Gen AI can interpret unstructured text and convert it into standardized systems, enabling large-scale data standardization without expert-driven manual efforts (Li et al. 2024). This capability allows unstructured information previously unconsidered in existing studies to be effectively represented and integrated into ML models. Accordingly, this study proposes a collaborative framework that leverages Gen AI to standardize OI and CI and integrates the standardized outputs into an ICDT prediction model. Through this framework, the impact of adopting Gen AI on both prediction performance and terminal productivity is empirically evaluated. In addition, the predicted results are applied to container stacking strategies, and extensive experiments are conducted to verify actual reductions in the number of yard crane relocations. The main contributions of this study are threefold. First, to the best of our knowledge, this study is the first to use Gen AI for standardizing unstructured data in the port logistics domain for the extraction of key ICDT determinants. Second, the standardized information is integrated into ML models, and state-of-the-art performance is demonstrated through experiments based on real container terminal data. Third, the effectiveness of the proposed framework is quantitatively validated by demonstrating reductions in the number of yard crane relocations achieved through ICDT-informed stacking strategies.

The remainder of this paper is organized as follows. Section 2 reviews related studies on ICDT prediction and AI-based standardization. Section 3 proposes a collaborative framework that integrates Gen AI and ML for ICDT prediction. Section 4 analyzes the relationship between Gen AI-based standardization results and ICDT. Section 5 evaluates the improvement in ICDT prediction performance achieved by incorporating standardized results into ML models. Section 6 compares the number of yard crane relocations through simulations that apply ICDT prediction to container stacking strategies. Section 7 provides an in-depth discussion of the technical and methodological implications and limitations of the proposed framework. Finally, Section 8 summarizes the research findings and presents directions for future research.

## 2. Literature Review

### 2.1 Applications of Generative AI in Maritime and Port Logistics

Artificial intelligence has been widely adopted in maritime and port logistics, with the majority of existing research being concentrated on machine learning and deep learning methods for prediction, optimization, and automation (Filom et al. 2022; Liu and Yuen 2025). These approaches have contributed to measurable improvements in operational efficiency, resource allocation, and in decision-making processes. Meanwhile, the emergence of Gen AI, whose capabilities extend beyond conventional predictive approaches, has attracted growing attention in these research areas (Richey et al. 2023; Jackson et al. 2024; Fosso Wamba et al. 2024). Jackson et al. (2024) proposed a capability-based framework for analyzing and implementing Gen AI in supply chain and operations management, while Fosso Wamba et al. (2024) examined the key benefits and challenges associated with its adoption in operations and supply chain management. In the maritime domain specifically, Reddy et al. (2024) proposed the application of Gen AI to maritime environments and discussed its potential to enhance safety and operational efficiency; however, the study also acknowledged that the maritime industry is still in the early stages of adopting this technology. Collectively, these studies suggest that the field is undergoing a transition from conventional machine learning approaches toward Gen AI; however, concrete applications in port and maritime operations have only recently begun to emerge (Liu et al. 2026; Badea et al. 2025).

Several recent studies have gone a step further by applying Gen AI to specific logistics problems. Wang et al. (2025) proposed STK-LLM, a berth recommendation framework that integrates spatiotemporal knowledge graphs with Gen AI agents, and reported meaningful operational improvements, including reduced vessel waiting times and increased berth utilization. Stavroulakis et al. (2026) proposed a multi-agent AI framework for coordinating freight handling, berth allocation, and fuel-efficiency decisions, thereby extending the applications of AI from independent optimization problems to a cross-functional negotiation architecture. It is significant because it demonstrates how Gen AI oriented intelligence can be embedded in dynamic port decision processes. In the maritime domain, Nguyen et al. (2025) introduced an open-source large language model specifically developed for maritime navigation, and demonstrated its superiority in critical decision-making tasks such as trajectory planning, risk assessment, and compliance with maritime regulations. Taken together, these studies indicate that the scope of Gen AI in these areas is expanding from conceptual discussions to more practical applications in forecasting, berth planning, and collaborative decision support.

These studies suggest several benefits of adopting Gen AI in maritime and port logistics. Hu and Li (2026) and Wang et al. (2025) both emphasized that it is essential to integrate Gen AI with conventional analytical methods, such as optimization models and knowledge graphs, rather than deploying it in isolation. Wang et al. (2025) further demonstrated that Gen AI could leverage embedded knowledge to reduce data labeling costs and support few-shot learning without the need for extensive domain-specific training data. Nguyen et al. (2025) noted that Gen AI could be extended beyond maritime navigation to other domains including logistics and environmental monitoring. However, several challenges remain to be addressed. Li et al. (2024) empirically confirmed that Gen AI was meaningful only when it was embedded into organizational practices and not when it was adopted as a standalone technology. Their finding implies that the operational value of Gen AI should be validated through concrete, domain-specific outcomes rather than assessing it solely at the level of technology adoption. Liu et al. (2026) identified interpretability, data quality, and domain adaptation as critical challenges constraining the reliable deployment of Gen AI in transportation systems, and Badea et al. (2025) reported that technical barriers and cost considerations remain major obstacles to Gen AI-driven port logistics. These studies suggest that integrating Gen AI with conventional methods can help enhance its practical applicability in domain-specific operational contexts.

### 2.2 Import Container Dwell Time Prediction

The primary obligation of container terminals is to ensure punctuality in the execution of vessel operations while strictly adhering to the estimated time of departure (Park et al. 2024). However, achieving this objective remains challenging in practice. Vessels frequently fail to depart on time owing to unpredictable factors such as port handling inefficiency and congestion (Chu et al. 2024), with 93.6% of shipping delays attributed to vessels waiting for berths and inefficiencies in loading and unloading processes (Wang et al. 2025). Such failures result in demurrage charges, diminished port credibility, and increased logistics costs (Filom et al. 2022). Because the yard area functions as a buffer connecting seaside and landside operations within the terminal, the operational performance of the yard influences the seaside processes (Heilig et al. 2020; Filom et al. 2022). In particular, vessel turnaround time (VTT), which is defined as the interval between berthing and departure following cargo handling, is an important performance indicator for evaluating overall port efficiency and throughput capacity (D. Park et al. 2026; Filom et al. 2022). Accordingly, improvements in yard operations, such as reducing the number of yard crane rehandling movements, contribute to shorter vessel turnaround times and to higher overall performance across both seaside and port-wide processes (De Armas Jacomino et al. 2021). In this context, ICDT prediction has emerged as one of the most prominent research topics in yard planning (Xie et al. 2025). Accurate prediction of ICDT enables terminal operators to develop effective stacking strategies by positioning containers such that unnecessary rehandling during retrieval is minimized (Maldonado et al. 2019). Reducing unproductive movements improves yard efficiency and the overall flow of terminal operations (De Armas Jacomino et al. 2021). Against this background, a growing body of research has focused on ICDT prediction to support yard planning and terminal operations.

In studies on ICDT prediction, unstructured data such as OI and CI are recognized as key factors affecting dwell time; however, systematic methodologies for incorporating them into ML models remain insufficient. Early studies primarily attempted to address this issue through manual categorization. Kourounioti et al. (2016) noted the lack of attention to ICDT prediction and argued that accurate ICDT estimation is essential for enabling yard planners to stack containers in a manner that facilitates the retrieval of those with a high pickup probability. To this end, they converted unstructured CI into eight categories (e.g., edible, sensitive edible, chemicals, and plastic) and applied an artificial neural network, demonstrating that the categorized CI contributes to improved ICDT prediction performance. Similarly, Aminatou et al. (2018) proposed normalizing CI into three categories based on value density (monetary value per kilogram) and reported that cargo type and value density are closely related to ICDT. However, the generalizability of this approach remains limited, as the categorization relied on subjective, value-based criteria. De Armas Jacomino et al. (2021) analyzed and incorporated product information, which is similar to CI, into ICDT prediction models. However, the practical deployability of their approach was unclear, as it was not explicitly stated whether the information was preprocessed by experts or could be automatically obtained in advance. Subsequent studies expanded the set of considered variables and sought to improve model interpretability. Lee et al. (2024) incorporated external factors, such as strikes and weather conditions, that had not been considered previously, to reduce uncertainty in ICDT prediction and applied explainable AI techniques to analyze the influence of each of the variables. Although this study broadened the variable scope, it acknowledged that directly utilizing OI and CI may be more important and thus did not fully resolve the fundamental challenge. Akodia et al. (2024) included proxy variables that indirectly represent CI and OI, such as the trucking company and container value density. However, transforming these variables using one-hot encoding into high-dimensional vectors led to an excessive increase in the number of categories, resulting in overfitting and increased computational complexity. A common limitation across these studies is the semantic loss and limited generalizability that arise when converting unstructured data into structured representations. Overcoming these limitations requires transforming unstructured information into a standard classification system that can preserve its meaning effectively while maintaining a manageable number of categories.

Against this background, recent studies have attempted to automatically convert unstructured CI into standard codes using AI techniques. Xie et al. (2025) proposed extracting semantic information from CI using GloVe word embeddings and matching it to the Standard International Trade Classification code. They reported up to 6% improvement in predictive accuracy for a binary classification task that determines yard locations in an intermodal terminal, demonstrating the potential benefits of standardizing unstructured data. However, word-embedding-based approaches cannot distinguish contextual meanings of the same word or handle out-of-vocabulary terms, abbreviations, and technical jargon specific to the port logistics domain that are absent from the training data. Addressing these issues requires building and continuously updating a domain-specific vocabulary, which demands substantial resources and domain expertise. Consequently, while the work by Xie et al. (2025) represents a pioneering attempt to automatically extract semantic information from unstructured data, it suggests that more advanced AI techniques are necessary to comprehensively handle the non-standardized notation practices and heterogeneous rules observed in real-world operational data.

Methodological improvements in this direction are practically important, as improvements in ICDT prediction performance can directly translate into enhanced terminal operations. Accurate ICDT prediction provides essential information for establishing container stacking strategies and can effectively reduce the number of relocations required during terminal operations. Gaete et al. (2017) proposed a stacking strategy that assigns containers with similar dwell-time groups to adjacent locations based on ICDT prediction results, demonstrating a reduction in inefficient relocations. Maldonado et al. (2019) developed a heuristic methodology that determines container stacking positions based on predicted ICDT, empirically showing that this strategy can significantly reduce relocations in real container terminals. De Armas Jacomino et al. (2021) formulated ICDT prediction as an ordinal regression problem and reported that this approach not only improves prediction performance but also maximizes relocation reduction when applied to stacking strategies. Therefore, improving ICDT prediction performance by effectively handling unstructured information can directly contribute to enhancing terminal operational efficiency.

### 2.3 Standardization of CI and OI

Due to the unstructured nature of CI and OI, many studies have attempted to standardize them using ML and AI techniques. CI is a text-based data type that describes imported items and their contents, and it is recorded in various forms depending on the data source, such as cargo descriptions and trade declarations. He et al. (2021) proposed an ML framework to automate harmonized system (HS) code standardization in the customs trade declaration process. In that study, bidirectional encoder representations from transformers (BERT) was used to extract semantic features from text and classify the HS codes. However, they reported that standardization accuracy degrades substantially depending on data quality. Zhou et al. (2022) pointed out that structured product description is often ambiguous and lacks continuous context, which complicates classification, and proposed a framework based on convolutional neural networks. Sasana et al. (2025) compared traditional ML-, deep learning (DL)-, and natural language processing-based methods for HS code standardization. They identified several persistent challenges that could lead to substantial errors, including short and ambiguous text descriptions, severe class imbalance, and formatting issues. These studies commonly emphasized that classification performance is highly sensitive to data quality and that enhanced data preprocessing and robust handling of linguistic variations are essential for accurate classification. To overcome these limitations, recent efforts have introduced large language models (LLMs) for HS code classification (Marra De Artiñano et al. 2023). This study showed that LLMs can outperform conventional ML models without complex data preprocessing. However, its generalizability is limited, as it was evaluated only within the agriculture, fisheries, and food domain.

In contrast, OI is typically recorded as a few words representing the name of the importing entity, which provides limited information about the firm's industry sector or business activities. Consequently, prior studies have attempted standardization by integrating additional information rather than directly using OI. Wood et al. (2017) extracted textual data from companies' official websites and developed a model to classify north american industry classification system (NAICS) codes using term frequency–inverse document frequency (TF-IDF) and a multilayer perceptron. They reported severe noise in the training data and extreme imbalance in the amount of data across industries as key limitations. Oehlert et al. (2022) performed feature extraction by applying TF-IDF to business activity descriptions and tax filings and classified NAICS codes using a random forest model. They noted that their approach was suitable only for specific time periods and administrative datasets, resulting in limited generalizability, and that the characteristics of ML models make it difficult to identify a clear reasoning process behind the standardization results. To address these limitations, recent studies have introduced language models that can capture linguistic characteristics. For instance, Jagrič and Herman (2024) employed BERT to classify business descriptions into industry categories, demonstrating the potential of language models to perform industry classification by understanding textual semantics. However, their approach required extensive prior data collection and did not consider re-training issues arising from linguistic variation or generalizability to other data sources.

Existing studies that aim to standardize CI and OI share several common limitations. First, because they adopt supervised learning, they require the construction and maintenance of large-scale pre-labeled training datasets. Second, the resulting models often exhibit limited generalizability, as they are effective primarily for data originating from the same source or domain as the training data. Third, the lack of interpretability in standardization outputs prevents post-hoc handling. These limitations significantly reduce deployability in container terminal environments, where standardization must be performed immediately when an import container is unloaded, without additional resources or personnel, and where the reasoning behind the outputs must be verifiable. From this perspective, Gen AI represents a suitable alternative for container terminal environments. Leveraging extensive pre-trained knowledge, Gen AI can perform standardization directly from given inputs without complex preprocessing or the construction of training datasets, thereby offering high generalizability. Moreover, when information is insufficient, it can perform web searches in a human-like manner to augment the information required for reasoning, even from limited short text. Furthermore, by providing the reasoning behind the standardization outputs, Gen AI enables interpretation and response. Building on this review, this study standardizes OI and CI using Gen AI and integrates the standardized information into an ICDT prediction framework.

## 3. Collaborative Framework for ICDT Prediction

This section introduces a collaborative framework that standardizes CI and OI using Gen AI and integrates the standardized results with an ML model to predict ICDT. The overall framework is illustrated in Fig. 1. The table of abbreviations used in this study is provided in Appendix 3.

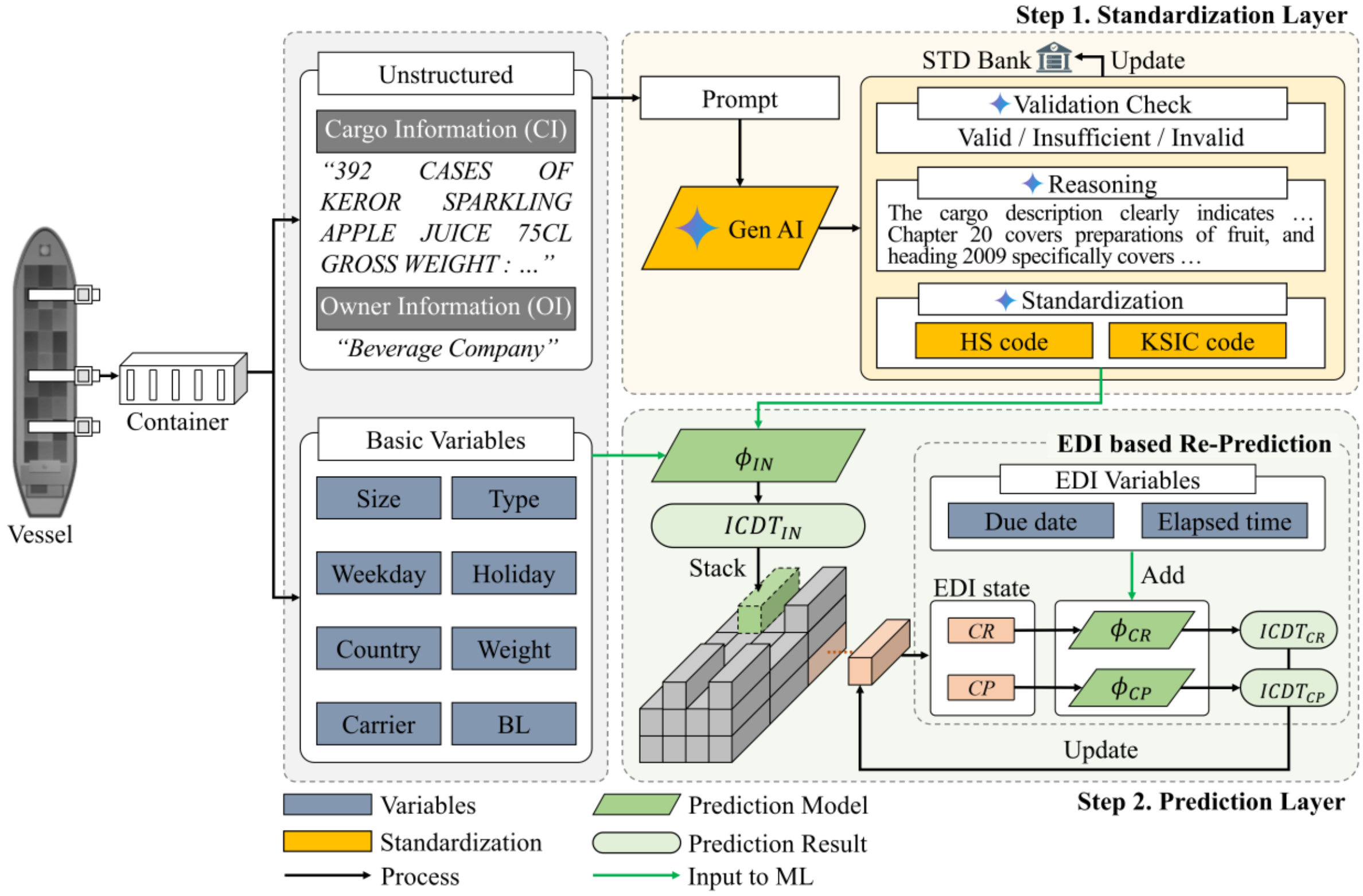

**Figure 1. Collaborative Framework of Generative AI and Machine Learning**

### 3.1 Standardization Layer

#### 3.1.1 HS and KSIC codes

The proposed framework adopts the following two codes for CI standardization.

○ **HS code**: The HS code is an international commodity classification system developed by the world customs organization and is currently used by more than 200 countries as the basis for tariff imposition and trade statistics (World Customs Organization, 1988). It provides a six-digit base structure, which individual countries further subdivide according to their tariff policies and statistical requirements. In this study, the two-digit hierarchical structure is defined as Standardization Levels 1–3 and is standardized using Gen AI.

○ **Korean Standard Industrial Classification (KSIC code)**: For OI standardization, the KSIC code is adopted. It is a national standard classification system established by Statistics Korea to systematically categorize economic activities based on the similarity of industrial activities. It is designed based on the United Nations' International Standard Industrial Classification while reflecting the industrial structure of Korea, and is used for economic policy formulation and industrial statistics (Statistics Korea, 2017). Because the focus of this study is the container terminal at Busan port, the KSIC code is adopted. However, depending on the geographical location of the container terminal, it can be replaced with the most appropriate industrial classification system. The KSIC code is hierarchically structured into one alphabetic character, two-digit numeric codes, and three-digit numeric codes, and is standardized into Levels 1–3 in the same manner as the HS code.

#### 3.1.2 Prompt Design

The performance of Gen AI varies depending on prompt design, and appropriately formulated instructions or task descriptions can yield high-quality results (Song et al. 2025; Zhang et al. 2025). The prompts for standardization in this study are designed with reference to these studies and are presented in Appendices 1 and 2.

A description of the standardization outputs is provided in Table 1. Gemini 2.5 Flash (Comanici et al. 2025) is adopted as the Gen AI model, considering its capability for large-scale batch processing and cost efficiency.

**Table 1. JSON Output Schema for HS/KSIC code Standardization**

| Category | Content |
|---|---|
| Standard Code (Level-1) | High-level Standard Code (HS/KSIC) |
| Standard Code (Level-2) | Middle-level Standard Code (HS/KSIC) |
| Standard Code (Level-3) | Low-level Standard Code (HS/KSIC) |
| Validation Check | Validation Check of Standardization: Type-1 (valid), Type-2 (insufficient), Type-3 (invalid) |
| Reason | Explanation for the standardized result |
| Evidence Tokens (only CI) | Key keywords or phrases that influenced the HS classification (e.g., "cotton 100%", "air conditioner parts") |
| Size (only OI) | Small and Medium Enterprise (SME), Mid, Large, or Unknown |

Table 1 summarizes the output fields produced by the Gen AI module when CI and OI are provided as inputs. Both CI and OI are standardized by Gen AI into a common hierarchical structure with three levels (Levels 1–3). In addition, to enable Gen AI to assess whether the input information is sufficient for standardization, the *Validation Check* field returns Type-1 (valid) when standardization is feasible, Type-2 (insufficient) when the information is inadequate, and Type-3 (invalid) when standardization is not possible. The basis for this assessment is briefly described in the *Reason* field. For CI, an additional *Evidence Tokens* field records the keywords that play a decisive role in the standardization process. This design enhances the explainability of the Gen AI–based standardization and enables the identification of improperly recorded data, thereby facilitating post hoc actions through discussion with domain experts.

In contrast, OI is often recorded as short text consisting of one to three words, such as company names, which makes semantic interpretation difficult. This characteristic poses a significant challenge for standardizing OI using Gen AI alone. Accordingly, when Gen AI determines that the given OI cannot be sufficiently identified based solely on its pre-trained knowledge, it is instructed to collect additional information through web search. Furthermore, since ICDT may vary with the size of the owning firm, even for the same cargo, a *Size* field is included to classify owner size for subsequent analysis.

### 3.1.3 Standardization Bank (STD Bank)

Since the use of Gen AI incurs monetary costs, improving the reuse rate of previously standardized information is essential for cost-efficient deployment. To this end, a standardization bank (STD Bank) is designed to store standardized results and reuse them when identical input information reappears, thereby minimizing repeated Gen AI calls. The operational mechanism of the STD Bank is presented in Algorithm 1 in the form of pseudocode. Although the algorithm is described with respect to CI, the same structure is applied to OI.

**Algorithm 1. Update Mechanism of Standardization Bank**

1: **INPUT**: $CI_{raw}$: Raw Cargo Information Text

2: **OUTPUT**: $CI_{std}$: Standardized result of $CI_{raw}$ (JSON format)

3: **DEFINITION**<br>
4: $STDBank_{CI}$: Standardization Bank of $CI_{raw}$<br>
5: $F_{std}(\cdot)$: GenAI − based standardization function<br>
6: **IF** $CI_{raw} \in STDBank_{CI}$ **THEN**<br>
7: | $CI_{std} \leftarrow STDBank_{CI}[CI_{raw}]$ # re-utilize standardized result<br>
8: | **RETURN** $CI_{std}$<br>
9: **ELSE**<br>
10: | $prompt_{CI_{raw}} \leftarrow Instruction_{CI} + TaskDescription_{CI} + CI_{raw}$ # request Gen AI<br>
11: | $CI_{std} \leftarrow F_{std}(prompt_{CI_{raw}})$<br>
12: | $STDBank_{CI}[CI_{raw}] \leftarrow CI_{std}$ # update STD Bank<br>
13: | **RETURN** $CI_{std}$<br>
14: **END IF**

Let $CI_{raw}$ denote the raw CI contained in an import container. When a $CI_{raw}$ is identified, the STD Bank is first queried to determine whether a standardized result for the corresponding information already exists. If no such result is found, a prompt $prompt_{CI_{raw}}$ is generated and provided as input to the Gen AI module, as described in Appendices 1 and 2. The standardized output returned by Gen AI is then updated in the STD Bank and defined as $CI_{std}$. Subsequently, when the same $CI_{raw}$ is encountered repeatedly, the previously standardized result $CI_{std}$ is retrieved directly from the STD Bank and returned without invoking Gen AI. Through this caching mechanism, redundant requests for identical $CI_{raw}$ are avoided, and as the number of stored $CI_{raw}$ entries in the STD Bank increases, the frequency of Gen AI calls is progressively reduced.

### 3.2 Prediction Layer

This subsection describes the input variables used in the prediction model. It also explains the process of redefining ICDT in response to the occurrence of electronic data interchange (EDI) events and predicting the updated ICDT.

#### 3.2.1 Basic Variables

When an import container is unloaded from a vessel, various types of information become available. Among them, the variables that can be directly input into the ML model are presented in Eq. (1).

$$X_{base} = \{c_{size}, c_{type}, c_{BL}, c_{country}, c_{carrier}, c_{day}, c_{holiday}\} \quad (1)$$

These variables represent container attributes such as size and type, as well as temporal and contextual information, including country, shipping line, day, and holiday. These variables are consistent with those commonly adopted in prior studies and are defined in this study as $X_{base}$.

#### 3.2.2 EDI-based Variables

After an import container is unloaded, multiple procedures must be completed before delivery. Tracking the progress of these procedures plays an important role in reducing uncertainty in ICDT prediction (Park et al. 2025). EDI is an electronic document exchange system that facilitates transportation flows by tracking various status changes occurring during the customs clearance process of import containers (Lee et al. 2000). In this study, through discussions with domain experts, EDI events that are closely related to ICDT are defined as *EDI states*, which are summarized in Table 2.

**Table 2. Description of EDI States**

| EDI Name | EDI State | Content |
| --- | --- | --- |
| Unloading | IN | **Container** is unloaded from the vessel and stacked in the storage yard |
| Container Release Order | CR | **Owner** completes customs procedures and obtains a release order |
| Container Pre-Information Notice | CP | **Carrier** submits pickup job information to the container terminal |
| Gate-out | OUT | **Container** is out of the gates and transported to the owner |

The *IN-state* represents the initial stage in which a container has been unloaded from a vessel, but no EDI message has yet been transmitted. Since customs-related procedures have not been completed at this stage, no EDI-based variables are generated. The *CR-state* indicates that customs clearance has been completed following the cargo owner's fulfillment of procedures such as duty payment and submission of required documents, and a delivery order (DO) is issued. The DO specifies the delivery due date by which the container must be released to the owner, denoted as $t_{do}$. At this stage, the elapsed time since unloading is defined as $ElapsedTime_e$ (Eq. (2)), using the unloading time $t_{IN}$ and EDI state update time $t_e$. In addition, the due-date variable $DueDate_e$ is defined in Eq. (3) using the delivery due date $t_{do}$ specified in the DO and the EDI update time $t_e$, where the temporal relationship satisfies $t_{IN} < t_{CR} < t_{CP} < t_{do}$.

$$ElapsedTime_e = t_e - t_{IN}, e \in \{CR, CP\} \quad (2)$$

$$DueDate_e = t_{do} - t_e, e \in \{CR, CP\} \quad (3)$$

$$X_{EDI} = \{ElapsedTime_e, DueDate_e\} \quad (4)$$

The *CP-state* is updated when an inland delivery company transmits information in advance for a container scheduled for handling at the terminal. This state is typically valid for 24–48 hours and indicates that a pickup truck has been assigned for the import container. This EDI event serves as an indicator that the departure of the import container is imminent. Accordingly, in the *CP-state*, both $ElapsedTime_e$ and $DueDate_e$ are updated and provided as inputs to the prediction model. These variables are defined as $X_{EDI}$ in Eq. (4).

### 3.2.3 Variables Generated from Gen AI

CI and OI are collected at the time of container unloading, in the same manner as $X_{base}$, but are represented as text. These variables are denoted as $X_{raw}$ in Eq. (5), and the Gen AI function for standardization is defined as $F_{std}$ in Eq. (6).

$$X_{raw} = \{c_{CI}, c_{OI}\} \tag{5}$$

$$F_{std}: X_{raw} \rightarrow X_{std} \tag{6}$$

Through $F_{std}$, $X_{raw}$ is standardized into $HS^*$,$KSIC^*$, as shown in Eqs. (7) and (8), and the standardized results are defined as $X^*_{std}$ (Eq. (9)).

$$HS^* = F_{std}(c_{CI}) \tag{7}$$

$$KSIC^* = F_{std}(c_{OI}) \tag{8}$$

$$X^*_{std} = \{HS^*, KSIC^*\} \tag{9}$$

The variables used for ICDT prediction are $X_{base}$, $X_{EDI}$, and $X^*_{std}$, and detailed descriptions of these variables are summarized in Table 3. The number of categories in $X^*_{std}$ varies depending on the level of standardization. Based on the experimental dataset used in this study, the most fine-grained standardization level, $HS^*_{lv3}$, contains 4,346 standardized codes, whereas $KSIC^*_{lv3}$ consists of 343 standardized codes. In Table 3, the column heading Source indicates the origin of the variable and Total denotes the number of unique values of the categorical variable. Each variable is used as an individual input to the prediction model, and the Examples provide the representative values. All variables, except ICDT, serve as input features for the prediction model; ICDT is the prediction target.

**Table 3. Input/Output Variables for ICDT Prediction**

| Variable | Type | Source | Total | Examples | Notes |
|---|---|---|---|---|---|
| $c_{size}$ | Category | Container | 2 | [20ft, 40ft] | Container size |
| $c_{type}$ | Category | Container | 4 | [Dry, Reefer, Danger, etc] | Container type |
| $c_{BL}$ | Category | Container | 4 | [0, 1, 2, None] | Bill of lading |
| $c_{country}$ | Category | Container | 44 | [CN, JP, ...] | Initial country of shipping |
| $c_{carrier}$ | Category | Container | 9 | [MSC, COSCO, ...] | Shipping line |
| $c_{day}$ | Category | Container | 7 | [Mon, Tue, ...] | Weekday at prediction time |
| $c_{holiday}$ | Binary | Container | 2 | [0, 1] | Holiday at prediction time |
| $ElapsedTime_e$ | Numeric | EDI | - | [12.5, 48.3, ...] | Time elapsed from *IN-state* to EDI state $e$ |
| $DueDate_e$ | Numeric | EDI | - | [24.0, 48.0, ...] | Remaining time to due date from EDI state $e$ |
| $HS^*_{lv1}$ | Category | Gen AI | 99 | [02, 04, ...] | |
| $HS^*_{lv2}$ | Category | Gen AI | 1,072 | [0202, 0402, ...] | HS code standardized by Gen AI |
| $HS^*_{lv3}$ | Category | Gen AI | 4,346 | [020230, 040240, ...] | |
| $KSIC^*_{lv1}$ | Category | Gen AI | 22 | [A, C, ...] | |
| $KSIC^*_{lv2}$ | Category | Gen AI | 75 | [A10, C14, ...] | KSIC code standardized by Gen AI |
| $KSIC^*_{lv3}$ | Category | Gen AI | 343 | [A101, C141, ...] | |
| $ICDT$ (target) | Numeric | Container | - | [120.5, 72.3, ...] | Remaining dwell time from current EDI state to gate-out |

***: Generated from Gen AI**

### 3.2.4 ICDT Definition and Prediction

Import containers and EDI states are defined as $c$ and $e$, respectively, as shown in Eqs. (10) and (11). Since an EDI state identifies the current stage of customs clearance for an import container, these two entities are always managed as a pair.

$$c = unloaded\ import\ container \tag{10}$$

$$e \in \{IN, CR, CP, OUT\} \tag{11}$$

Conventional ICDT is defined as the time interval from unloading to departure. However, by incorporating EDI states, ICDT can be redefined as shown in Eq. (12). Here, $t_{OUT}$ denotes the time at which the container departs from the terminal and $t_e$ represents the time at which the EDI state is updated, satisfying the temporal order $t_{IN} < t_{CR} < t_{CP} < t_{OUT}$. The redefined $ICDT_c^e$ in Eq. (12) represents the remaining dwell time from the current EDI state to the departure time. The ML model used for ICDT prediction is denoted by $\phi$, and separate models are trained for each EDI state to predict $ICDT_c^e$ (Eqs. 13–15).

$$ICDT_c^e = t_{OUT} - t_e, e \in \{IN, CR, CP\} \tag{12}$$

$$\phi_{IN}: \{X_{base}, X_{std}^*\} \rightarrow ICDT_c^{IN} \tag{13}$$

$$\phi_{CR}: \{X_{base}, X_{EDI}, X_{std}^*\} \rightarrow ICDT_c^{CR} \tag{14}$$

$$\phi_{CP}: \{X_{base}, X_{EDI}, X_{std}^*\} \rightarrow ICDT_c^{CP} \tag{15}$$

$ICDT_c^{IN}$ in Eq. (16) represents the ICDT predicted by $\phi_{IN}$ at the moment when an import container is unloaded (i.e., at time $t_{IN}$). At this stage, since no EDI state has been updated, $X_{EDI}$ is not utilized. The predicted $ICDT_c^{IN}$ is regarded as the current ICDT of the container, denoted by $ICDT_c$, until an EDI state update occurs (Eq. 17).

$$ICDT_c^{IN} = \phi_{IN}(X_{base}, X_{std}^*) \tag{16}$$

$$ICDT_c \leftarrow ICDT_c^{IN} \tag{17}$$

At time $t_e$, when an EDI state is updated (i.e., *CR* or *CP*), the corresponding container is already located in the storage yard, and only its operational status changes. As shown in Eq. (18), $\phi_e$ re-predicts $ICDT_c^e$ and updates $ICDT_c$ (Eq. 19). The rationale for re-predicting ICDT according to EDI states is to effectively capture changes in the ICDT distribution associated with state transitions. This process terminates once the container departs from the terminal.

$$ICDT_c^e = \phi_e(X_{base}, X_{std}^*, X_{EDI}), e \in \{CR, CP\} \tag{18}$$

$$ICDT_c \leftarrow ICDT_c^e \tag{19}$$

## 4. Gen AI–Based Standardization of OI and CI

### 4.1 Illustrative Examples of Standardization

This subsection presents example cases of the standardization results produced by Gen AI. Table 4 illustrates cases in which CI is standardized to the Level-2 HS code "2009."

In Example 1, the CI is recorded as "MIX VEGETABLE CONCENTRATE ORDER," and the Gen AI identifies "VEGETABLE" and "CONCENTRATE" as the key tokens. In Example 2, "KEROR," "SPARKLING," "APPLE," and "JUICE" are identified as the salient terms, whereas in Example 3, "POMEGRANATE" and "CONC." are extracted as the primary tokens. Notably, even when identical meanings are expressed using different notations, such as "CONCENTRATE" in Example 1 and "CONC." in Example 3, Gen AI successfully identifies them as key tokens. In addition, "KEROR" in Example 2 refers to a fruit juice brand that is generally difficult to recognize without prior knowledge; nevertheless, Gen AI identifies it as a key token. These results demonstrate that Gen AI can effectively recognize and standardize recorded information despite inconsistencies in notation and the presence of domain-specific or uncommon terms.

**Table 4. Example of HS code Standardization Based on CI**

| HS code | Description | Standardization Results | |
|---|---|---|---|
| 2009 | Fruit or nut juices (including grape must and coconut water) and vegetable juices, unfermented and not containing added spirit, whether or not containing added sugar or other sweetening matter | Example 1 | MIX VEGETABLE CONCENTRATE ORDER |
| | | Evidence Tokens* | [“VEGETABLE”, “CONCENTRATE”] |
| | | Reason* | The cargo description “MIX VEGETABLE CONCENTRATE” clearly indicates a processed vegetable product. The term “CONCENTRATE” is generally classified as a juice form, while “VEGETABLE” refers to vegetable-based items. HS Code 2009 covers non-fermented vegetable juices, and “MIX” corresponds to mixed juices, which are most appropriately classified under subheading 200990. |
| | | Example 2 | 392 CASES OF KEROR SPARKLING APPLE JUICE 75CL GROSS WEIGHT : 6664 KGS, SPARKLING WHITE GRAPE JUICE 75CL GROSS WEIGHT : 8568 KGS |
| | | Evidence Tokens* | [“KEROR”, “SPARKLING”, “APPLE”, “GRAPE”, “JUICE”] |
| | | Reason* | The cargo description clearly indicates “SPARKLING APPLE JUICE” and “SPARKLING WHITE GRAPE JUICE”, which are unfermented fruit juices. Chapter 20 covers preparations of fruit, and heading 2009 specifically covers unfermented fruit juices (including grape must). Subheading 200979 for “other apple juice” is chosen as the representative HS Code for these unfermented, sparkling fruit juices. The presence of two distinct types of juice confirms multiple items. |
| | | Example 3 | 80 DRUMS OF CLEAR POMEGRANATE CONC. VARIETY BLEND8319 80 DRUMS OF CLEAR POMEGRANATE CONC. |
| | | Evidence Tokens* | [“POMEGRANATE”, “CONC.”] |
| | | Reason* | The item is identified as 'POMEGRANATE CONC.', referring to pomegranate concentrate. This product, being an unfermented fruit juice preparation, is classified under Chapter 20 for preparations of fruit. Specifically, Heading 2009 covers fruit juices, and Subheading 200989 is appropriate for other fruit or vegetable juices not specified elsewhere. |

*** Generated from Gen AI**

**Table 5. Example of KSIC code Standardization Based on OI**

| KSIC | Description | Results | |
|---|---|---|---|
| I561 | Restaurants and mobile food service activities | Example 1 | Company-4 |
| | | Reason* | The company “Company-4” primarily engages in food manufacturing and distribution, as well as restaurant and contract catering businesses. **Multiple information sources** specify its detailed business category as “Institutional Cafeteria Services.” According to the Korean Standard Industrial Classification (KSIC), this corresponds to the major category I. Accommodation and Food Service Activities, the middle category 56. Food and Beverage Service Activities, and the subcategory 561. Restaurants and Mobile Food Service Activities. |
| | | Example 2 | Company-5 |
| | | Reason* | The company “Company-5” is a food service subsidiary, operating a wide range of restaurants including buffets, dining establishments, and bakeries. This corresponds to the major category I. Accommodation and Food Service Activities, the middle category 56. Food and Beverage Service Activities, and the subcategory 561. Restaurants and Mobile Food Service Activities. |
| | | Example 3 | Company-6 |
| | | Reason* | The company “Company-6” is identified through **web search results as** “X-6 Korea”, a company providing in-flight catering services. Catering Services, which falls under the major category I. Accommodation and Food Service Activities, the middle category 56. Food and Beverage Service Activities, and the subcategory 561. Restaurants and Mobile Food Service Activities. |

*** Generated from Gen AI**

Table 5 presents cases in which OI is classified into the Level-3 KSIC code “I561,” which corresponds to *Restaurants and mobile food service activities*. During this process, when information regarding the owner is insufficient, Gen AI attempts to supplement the information through web search, as indicated by outputs such as “multiple information sources …” in Example 1 and “web search results as …” in Example 3. This behavior indicates that Gen AI autonomously assesses the adequacy of OI and explores additional information sources when necessary. These examples demonstrate that Gen AI can collect information and perform standardization through a process analogous to human reasoning, without requiring explicit data preprocessing or manual data collection.

### 4.2 Validation of Standardization Results

Although this study instructs Gen AI to perform standardization, not all OI and CI can be successfully standardized. OI and CI are generated by multiple stakeholders and may be incomplete due to heterogeneous recording practices. Therefore, identifying such data in advance is essential for improving data quality. In this subsection, the validation mechanism of Gen AI, as defined in Section 3.1.2, is evaluated by analyzing hierarchical standardization results to assess whether Gen AI can autonomously determine the feasibility of standardization based on the given input information.

**Table 6. Standardization Validation by Gen AI Assessment and Standardization Level**

| Category | Validation Check* | # of Standardization (a) | Level* | Non-matched (b) | Ratio (b/a) |
|---|---|---|---|---|---|
| KSIC code | Type-1 (valid) | 22,271 (93.05%) | Lv. 1 | 1 | 0.00% |
| | | | Lv. 2 | 3 | 0.00% |
| | | | Lv. 3 | 921 | 4.14% |
| | Type-2 (insufficient) | 669 (2.80%) | Lv. 1 | 0 | 0.00% |
| | | | Lv. 2 | 0 | 0.00% |
| | | | Lv. 3 | 31 | 4.63% |
| | Type-3 (invalid) | 994 (4.15%) | Lv. 1 | 945 | 95.07% |
| | | | Lv. 2 | 946 | 95.17% |
| | | | Lv. 3 | 950 | 95.57% |
| | Total | 23,934 | | | |
| HS Code | Type-1 (valid) | 106,775 (94.51%) | Lv. 1 | 0 | 0.00% |
| | | | Lv. 2 | 0 | 0.00% |
| | | | Lv. 3 | 3,671 | 3.44% |
| | Type-2 (insufficient) | 6,190 (5.48%) | Lv. 1 | 4,748 | 76.70% |
| | | | Lv. 2 | 4,748 | 76.70% |
| | | | Lv. 3 | 5,898 | 95.28% |
| | Type-3 (invalid) | 5 (0.01%) | Lv. 1 | 4 | 80.00% |
| | | | Lv. 2 | 4 | 80.00% |
| | | | Lv. 3 | 4 | 80.00% |
| | Total | 112,970 | | | |

***: Generated from Gen AI**

Table 6 presents the validation results that compare the consistency between the standardization outputs generated by Gen AI and the actual standard code systems, categorized by *Validation Check* types. A case is counted as *non-matched* when the standardization result produced by Gen AI corresponds to an invalid standard code in the ground-truth classification system. Since identical OI and CI may appear across multiple containers, deduplication is performed, and the Gen AI standardization is executed only once.

Focusing first on the KSIC code standardization results, among the 22,271 cases classified as Type-1, only one case (0.004%) at Level-1 and three cases (0.013%) at Level-2 were identified as non-matched. At the most fine-grained level (Level-3), 921 cases (4.14%) were classified as non-matched. For Type-2, out of 669 cases, non-matched results were observed only at Level 3, amounting to 31 cases (4.63%). In contrast, for Type-3, approximately 95% of the 994 cases were identified as non-matched across all levels. This indicates that Gen AI can reliably identify cases that are not suitable for standardization.

A similar trend is observed in the HS code standardization results. Among the 106,775 Type-1 cases, non-matched results were observed only at Level-3, accounting for 3,671 cases (3.44%), while all cases at the Level-1 and Level-2 classifications were correctly matched. For Type-2, which consists of 6,190 cases, non-matched results accounted for 4,748 cases (76.70%) at both Level-1 and Level-2 and increased to 5,898 cases (95.28%) at Level-3. Type-3 cases were extremely rare, with only four instances observed, in contrast to the OI results. This difference can be attributed to the fact that CI, even when recorded in highly abbreviated form, typically contains some cargo-related information, which Gen AI recognizes as insufficient for standardization, leading to Type-2 classification rather than Type-3.

In summary, the low proportion of non-matched cases in Type-1 demonstrates that when Gen AI determines standardization to be feasible, it assigns valid standard codes. Conversely, the high non-matched ratios observed in Type-2 and Type-3 indicate that Gen AI accurately identifies cases with insufficient information or those unsuitable for standardization. These results demonstrate that the proposed self-validation mechanism can effectively determine the standardizability of input information prior to standardization.

### 4.3 ICDT Distribution by Standardized Categories

This subsection analyzes differences in ICDT according to the standardization results of KSIC and HS codes. For this purpose, four high-frequency categories at Level-1 are first selected, and up to three subcategories are extracted from each of them. For these categories, the proportion of observations and the empirical ICDT distributions are presented, together with the results of independent *t-tests* to indicate statistical significance. The overall results of this analysis are summarized in Fig. 2.

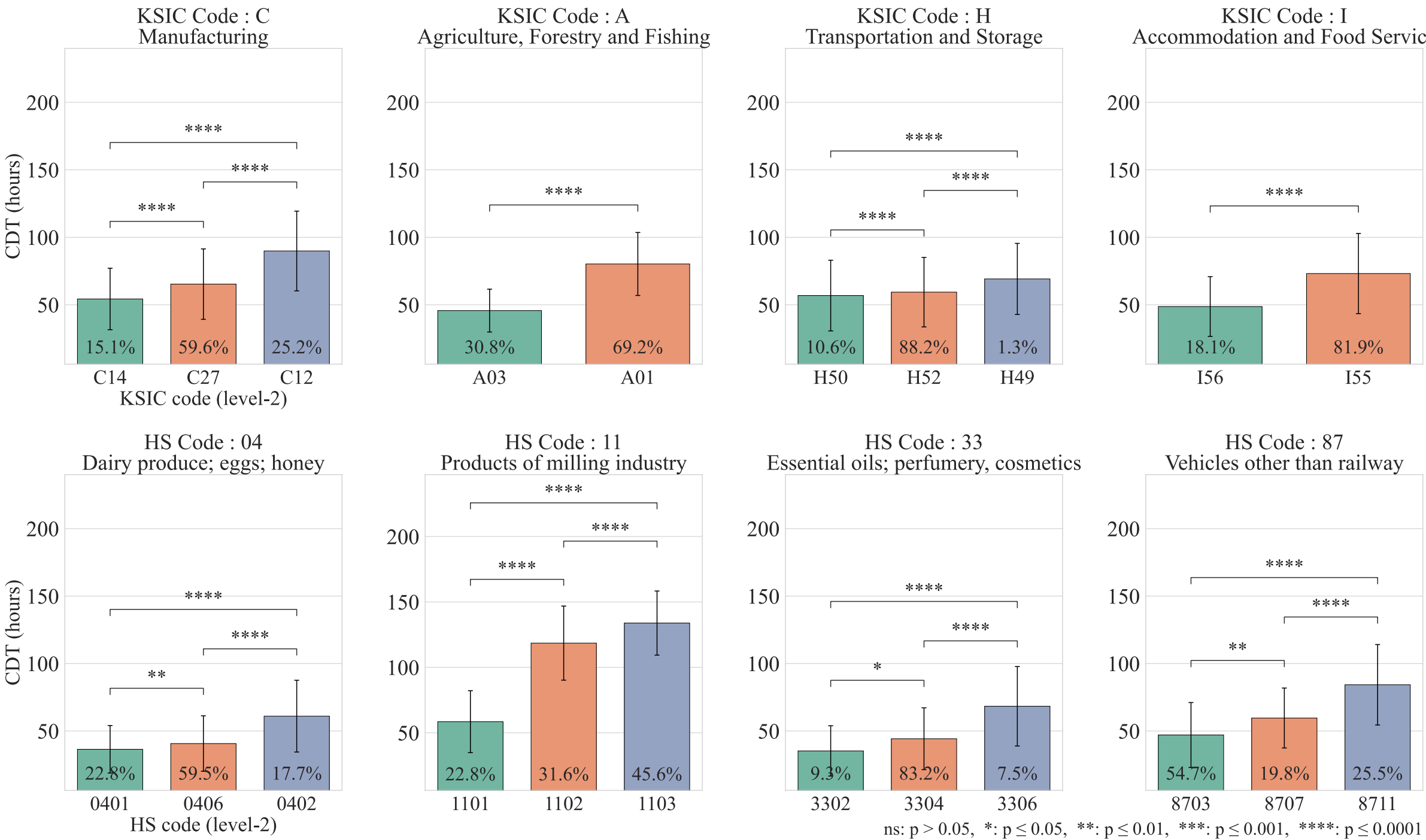

**Figure 2. ICDT Distribution Based on Standardization Result**

From the KSIC code analysis, the most frequently standardized Level-1 categories were C (Manufacturing), A (Agriculture, Forestry, and Fishing), H (Transportation and Storage), and I (Accommodation and Food Service). Within category C (Manufacturing), the identified subcategories were C14 (Manufacture of tobacco products), C27 (Manufacture of medical products), and C12 (Manufacture of wearing apparel), which exhibited ICDT values of 54.29, 65.30, and 89.86 h, respectively. Statistically significant differences were observed among these subcategories. Category A (Agriculture, Forestry, and Fishing) consisted of two subcategories: A03 (Fishing and aquaculture) and A01 (Agriculture), with corresponding ICDT values of 45.69 and 80.21 h. For category H (Transportation and Storage), three subcategories were identified: H50 (Water transport), H52 (Warehousing and support activities), and H49 (Land transport), whereas category I (Accommodation and Food Service) comprised two subcategories, I56 (Food and beverage service activities) and I55 (Accommodation). At Level-2, the KSIC code standardization results exhibited statistically significant differences across all the analyzed categories.

The HS code analysis revealed that the most frequently standardized Level-1 categories were 04 (Dairy produce; eggs; honey), 11 (Products of the milling industry), 33 (Essential oils, perfumery, cosmetics), and 87 (Vehicles other than railway). Within category 04 (Dairy produce; eggs; honey), the subcategories 0401 (Milk and cream, not concentrated nor containing added sugar), 0406 (Cheese and curd), and 0402 (Milk and cream, concentrated or containing added sugar) exhibited ICDT values of 36.45, 40.71, and 61.01 h, respectively. For category 11 (Products of the milling industry), the subcategories 1101 (Wheat or meslin flour), 1102 (Cereal flours other than of wheat or meslin), and 1103 (Cereal groats, meal, and pellets) showed ICDT values of 58.50, 118.53, and 133.84 h, respectively, indicating substantial differences in ICDT even within the same parent category. In category 33 (Essential oils, perfumery, cosmetics), relatively shorter ICDT values were observed, with 3302 (Mixtures of odoriferous substances) at 35.25 h, 3304 (Beauty or make-up preparations and preparations for the care of the skin) at 44.26 h, and 3306 (Preparations for oral or dental hygiene, including denture fixative pastes and powders) at 68.34 h. Finally, within category 87 (Vehicles other than railway), the subcategories 8703 (Motor cars and other motor vehicles principally designed for the transport of persons), 8707 (Bodies (including cabs), for motor vehicles), and 8711 (Motorcycles and cycles fitted with an auxiliary motor) exhibited ICDT values of 47.05, 59.62, and 84.27 h, respectively. All of these differences were statistically significant, demonstrating that Gen AI-standardized results constitute important variables that explain variations in ICDT.

### 4.4 Interactive Effects of Standardized OI and CI on ICDT

Although OI and CI are standardized independently by Gen AI, their combination may reveal additional heterogeneity in ICDT. To examine this hypothesis, this study analyzes how ICDT distributions vary within Level-2 HS codes according to owner size, as classified by Gen AI. For this purpose, the top 100 Level-2 HS codes with the largest sample sizes are identified, and representative codes from this set are selected for independent t-tests. The results are presented in Fig. 3. SMEs denote small and medium-sized enterprises as classified by Gen AI based on owner-related information.

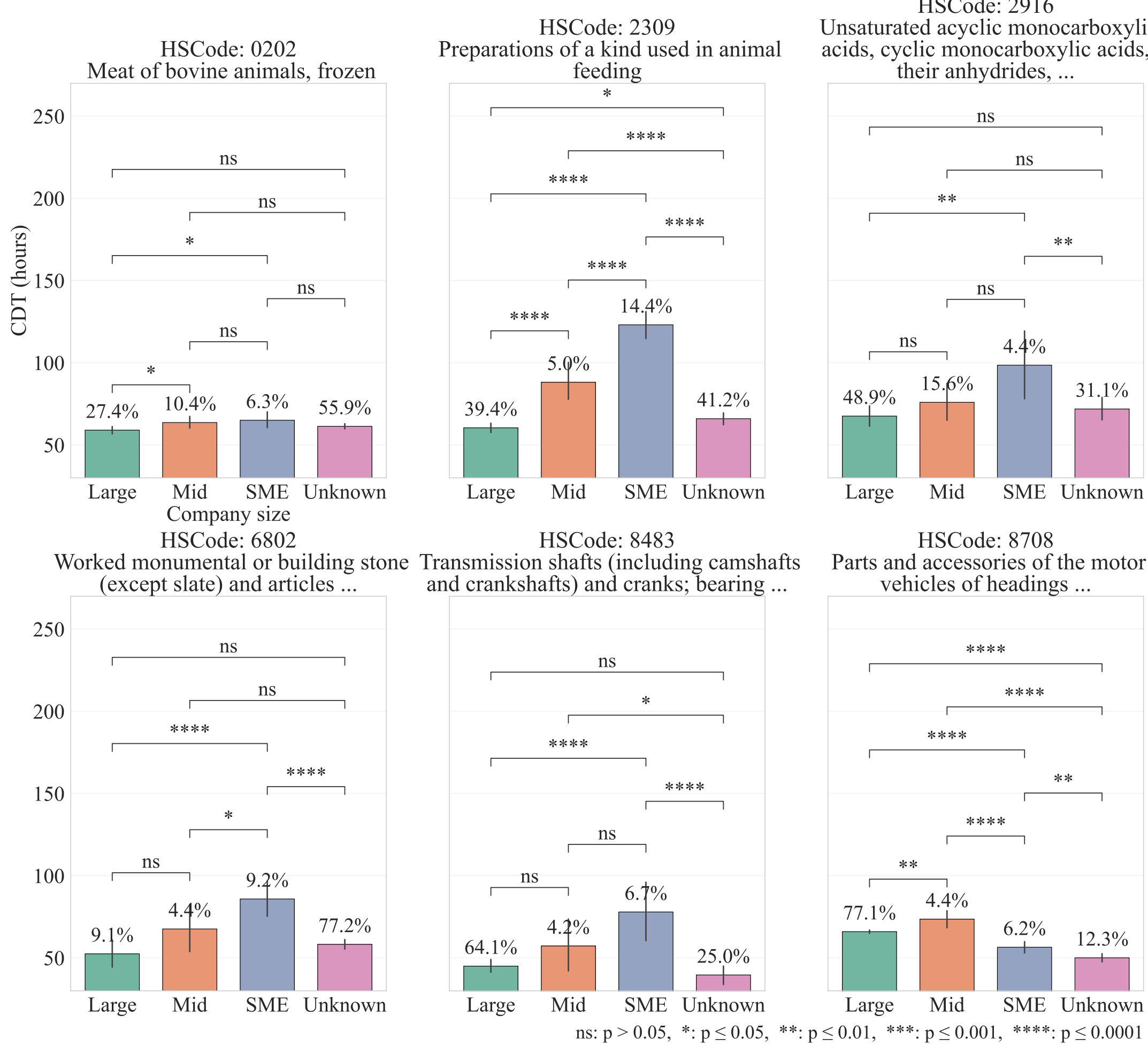

**Figure 3. ICDT Distribution Based on HS Code and Size Class**

For HS code 0202 (Frozen bovine meat), a large proportion of owners were classified as “Unknown”, and ICDT values appeared similar regardless of owner size. Although statistical significance was observed for a limited number of combinations, most comparisons did not exhibit significant differences. This suggests that for perishable goods such as frozen meat, rapid departure is essential due to product characteristics, thereby limiting the influence of owner size on ICDT. In contrast, for HS codes 2309 (Animal feed preparations), 2916 (Unsaturated acyclic monocarboxylic acids), 6802 (Worked monumental/building stone), and 8483 (Transmission shafts, gears, and bearings), a consistent pattern emerged in which larger owner size was associated with shorter ICDT and reduced variance in the distribution. In particular, comparisons between large firms and SMEs exhibited strong statistical significance across all cases. This pattern suggests that larger owners tend to operate standardized logistics processes and adhere to more clearly defined departure schedules. Conversely, SMEs exhibited greater dispersion in ICDT, likely due to their smaller representation and greater variability in departure timing across individual firms. By contrast, HS code 8708 (Motor vehicle parts) exhibited a different pattern, with SMEs showing shorter ICDT than large firms, deviating from the trends observed for the other product categories.

In summary, when product characteristics such as perishability primarily determine the speed of departure, the effect of owner size on the ICDT is limited. However, for certain product categories, statistically significant differences in ICDT by owner size are observed. These results demonstrate that OI and CI play important roles in explaining ICDT not only individually but also in combination, indicating that the two sources of information exhibit complementary characteristics.

## 5. ICDT Prediction Experiments

This section analyzes the impact of Gen AI–generated standardization results on the performance of ICDT prediction models and identifies models that achieve state-of-the-art performance.

### 5.1 Experimental Settings

The training dataset consists of two years of data collected from an operational container terminal at Busan port between 2024 and 2025, comprising 394,386 records. In consultation with domain experts, records with ICDT values exceeding 10 days are excluded from the training process. The variables used in the experiments are summarized in Section 3.2.3. Tree-based ML models that are commonly adopted in ICDT prediction studies, namely extreme gradient boosting (XGBoost), light gradient boosting machine (LightGBM), and categorical boosting (CatBoost), which is known for effectively modeling high-cardinality categorical variables, are selected for evaluation (Chen and Guestrin 2016; Ke et al. 2017; Prokhorenkova et al. 2018). These models have been widely reported to outperform DL models on tabular data over an extended period (Grinsztajn et al. 2022; Shwartz-Ziv and Armon 2022). To evaluate non-tree-based ML models, support vector regressor (SVR), which performs regression through kernel-based mapping in high-dimensional feature spaces, and multi-layer perceptron (MLP), a feedforward neural network that learns nonlinear relationships through multiple hidden layers, are also selected (Rumelhart et al. 1986, Drucker et al. 1997). In addition, DL models specialized for tabular data processing, including FT-Transformer (FTT), neural oblivious decision ensembles (NODE), and TabNet, are included for comparison (Arik and Pfister 2021; Gorishniy et al. 2021; Popov et al. 2020). Finally, a baseline model that predicts the mean ICDT value is adopted to benchmark the performance improvements achieved by the advanced models.

$$MAE = \frac{1}{n} \sum_{i}^{n} |\hat{y}_i - y_i| \tag{20}$$

To evaluate ICDT prediction performance, mean absolute error (MAE) is adopted, as it directly represents the absolute difference between predicted and observed values in units of time (Eq. 20). MAE has been widely used as an evaluation metric in prior ICDT prediction studies (De Armas Jacomino et al. 2021). For both KSIC and HS codes, Level-3 standardized results are used. To enhance the generalizability of the experimental results, the test dataset is randomly sampled at ratios of 20%, 30%, and 40% of the full dataset. Each experiment is repeated 10 times, and the mean and standard deviation of the performance metrics are reported.

### 5.2 Model Performance and Analysis

This subsection presents a comparative analysis of ICDT prediction performance with and without Gen AI-generated standardization results. Table 7 summarizes the results for two settings: (a) without standardization and (b) with standardization, evaluated across multiple benchmark models, three test ratios (20%, 30%, 40%), and three EDI states (*IN*, *CR*, *CP*). Bold values indicate the better performance between settings (a) and (b) for each model and EDI state, while underlined values denote the overall best-performing model for each EDI state. The following analysis focuses on the test ratio of 20%.

**Table 7. ICDT Prediction Performance Comparison (Iteration: 10)**

| Test Ratio | Model | (a) Without Standardization Result | | | (b) With Standardization Result | | |
|---|---|---|---|---|---|---|---|
| | | IN | CR | CP | IN | CR | CP |
| | MEAN | 81.464±0.464 | 38.706±0.200 | 2.503±0.006 | 81.464±0.464 | 38.706±0.200 | 2.503±0.006 |
| | XGBoost | 42.113±2.445 | 29.931±1.217 | **2.096±0.117** | **37.470±2.818** | **27.981±2.712** | 2.116±0.172 |
| | LightGBM | 38.652±0.105 | 27.214±0.080 | 2.067±0.004 | **34.144±0.528** | **24.243±0.757** | **1.999±0.053** |
| | CatBoost | 38.613±0.097 | 26.913±0.072 | 2.033±0.009 | **33.253±0.183** | **23.710±0.578** | **1.892±0.008** |
| 20% | SVR | 38.467±0.105 | 28.372±0.118 | 2.249±0.015 | **36.409±0.117** | **28.214±0.125** | **2.199±0.008** |
| | MLP | 38.057±0.159 | 28.052±0.106 | 2.197±0.009 | **35.212±0.123** | **26.193±0.137** | **2.034±0.013** |
| | FTT | 64.306±0.177 | **40.576±0.213** | 2.981±0.010 | **64.302±0.181** | 40.577±0.207 | **2.960±0.023** |
| | NODE | 64.298±0.168 | 40.578±0.209 | 2.982±0.005 | **64.296±0.162** | **40.572±0.203** | **2.980±0.009** |
| | TabNet | **62.859±0.967** | 40.499±0.233 | **2.746±0.206** | 63.398±1.031 | **40.362±0.321** | 2.793±0.174 |
| | MEAN | 81.361±0.211 | 38.643±0.103 | 2.507±0.004 | 81.361±0.211 | 38.643±0.103 | 2.507±0.004 |
| | XGBoost | 40.504±1.838 | 29.767±1.870 | **2.083±0.116** | **36.968±2.981** | **27.617±2.643** | 2.084±0.183 |
| | LightGBM | 38.647±0.075 | 27.211±0.073 | 2.066±0.003 | **34.175±0.519** | **24.308±0.744** | **2.001±0.050** |
| | CatBoost | 38.631±0.070 | 26.923±0.070 | 2.037±0.007 | **33.356±0.170** | **23.793±0.564** | **1.989±0.092** |
| 30% | SVR | 38.456±0.091 | 28.341±0.094 | 2.242±0.004 | **36.418±0.101** | **28.194±0.114** | **2.213±0.024** |
| | MLP | 38.134±0.117 | 28.038±0.114 | 2.199±0.009 | **35.499±0.151** | **26.298±0.122** | **2.039±0.007** |
| | FTT | 64.359±0.220 | 40.669±0.132 | **2.982±0.321** | **64.296±0.218** | **40.565±0.109** | 3.205±0.329 |
| | NODE | 64.322±0.187 | 40.658±0.150 | 3.016±0.021 | **64.311±0.215** | **40.628±0.147** | **2.994±0.016** |
| | TabNet | 64.315±0.197 | **40.585±0.100** | **2.973±0.011** | **64.312±0.186** | 40.634±0.135 | 3.158±0.254 |
| | MEAN | 81.346±0.208 | 38.627±0.088 | 2.506±0.001 | 81.346±0.208 | 38.627±0.088 | 2.506±0.001 |
| | XGBoost | 40.277±1.370 | 29.445±2.108 | **2.072±0.059** | **36.885±2.983** | **27.336±2.362** | 2.087±0.172 |
| | LightGBM | 38.640±0.055 | 27.228±0.060 | 2.067±0.005 | **34.237±0.537** | **24.393±0.745** | **2.005±0.049** |
| | CatBoost | 38.629±0.084 | 26.976±0.051 | 2.040±0.006 | **33.463±0.200** | **23.982±0.573** | **1.995±0.090** |
| 40% | SVR | 38.445±0.065 | 28.351±0.067 | 2.250±0.011 | **36.441±0.070** | **28.213±0.103** | **2.221±0.026** |
| | MLP | 38.144±0.118 | 28.107±0.110 | 2.197±0.008 | **35.771±0.096** | **26.506±0.094** | **2.053±0.007** |
| | FTT | **64.255±0.137** | **40.596±0.079** | 3.008±0.015 | 64.274±0.172 | 40.597±0.020 | **2.972±0.027** |
| | NODE | 64.512±0.117 | 40.876±0.099 | 3.013±0.011 | **64.265±0.215** | **40.745±0.054** | **2.824±0.011** |
| | TabNet | **64.261±0.176** | 40.624±0.093 | **3.023±0.077** | 64.278±0.185 | **40.622±0.050** | 3.028±0.031 |

In experiment (a), which does not utilize standardization results, the mean predictor reported MAE values of 81.464, 38.706, and 2.503 for the *IN*-, *CR*-, and *CP-states*, respectively, representing the lowest performance among all benchmarks. DL models including FTT, NODE, and TabNet exhibited consistently inferior performance across all test ratios and EDI states. In particular, for the *CP-state*, the average MAE of the DL models reached 2.903, corresponding to a 15.98% performance degradation compared to the mean-value baseline. This observation is consistent with prior studies that have reported the limitations of DL models in extracting meaningful information from tabular data (Grinsztajn et al. 2022; Shwartz-Ziv and Armon 2022). In contrast, tree-based ML models demonstrated substantially better performance. XGBoost achieved MAE values of 42.113, 29.931, and 2.096 for the *IN*-, *CR*-, and *CP-states*, respectively, while LightGBM reported 38.652, 27.214, and 2.067. CatBoost achieved MAE values of 38.613, 26.913, and 2.033, showing the best performance in the *CR*- and *CP-states*. SVR and MLP, which are not tree-based models, achieved MAE values of 38.467, 28.372, 2.249 and 38.057, 28.052, 2.197, respectively, with MLP showing the best performance in the *IN-state*.

In experiment (b), which incorporates the standardization results, DL models did not exhibit noticeable performance improvements compared to experiment (a). In contrast, the ML models showed consistent performance gains in most cases. For XGBoost, MAE values improved from 42.113, 29.931, and 2.096 in experiment (a) to 37.470, 27.981, and 2.116 in experiment (b), corresponding to performance improvements of 11.03% and 6.51% for the *IN*- and *CR-states*, respectively, although a slight degradation of 0.95% was observed for the *CP-state*. LightGBM improved from 38.652, 27.214, and 2.067 to 34.144, 24.243, and 1.999, achieving performance gains of 11.66%, 10.92%, and 3.29% across all EDI states. CatBoost achieved the largest improvements, with MAE values decreasing from 38.613, 26.913, and 2.033 in experiment (a) to 33.253, 23.710, and 1.892 in experiment (b), corresponding to improvements of 13.88%, 11.90%, and 6.94%, respectively. SVR also showed performance improvements across all the EDI states, with MAE values decreasing from 38.467, 28.372, and 2.249 to 36.409, 28.214, and 2.199, respectively. However, compared to the tree-based models, the improvement was limited because kernel-based distance computation becomes less effective in the sparse, high-dimensional feature space created by the categorical variables. MLP improved from 38.057, 28.052, and 2.197 to 35.212, 26.193, and 2.034, respectively, achieving performance gains of 7.48%, 6.63%, and 7.42% across all the EDI states. DL models, including FTT, NODE, and TabNet, did not exhibit consistent performance improvements across test ratios. Nevertheless, the substantial improvements achieved by tree-based models confirm that the standardized information effectively enhances prediction performance.

In addition, the effect of the standardization level of OI and CI on ICDT prediction performance is analyzed. To isolate the impact of standardization level, CatBoost, which achieved the best performance across the previous experiments, is trained for nine combinations of KSIC and HS codes. For each combination, experiments are repeated ten times with the test ratio fixed at 20%. The results are summarized in Table 8.

**Table 8. Prediction Performance by KSIC and HS code Standardization Level**

| EDI State | HS code | KSIC code Lv. 1 | KSIC code Lv. 2 | KSIC code Lv. 3 | Avg |
|---|---|---|---|---|---|
| IN | Lv. 1 | 36.344±0.118 | 35.850±0.163 | 35.241±0.140 | 35.812±0.451 |
| | Lv. 2 | 34.878±0.108 | 34.405±0.212 | 33.894±0.222 | 34.392±0.402 |
| | Lv. 3 | 34.222±0.134 | 33.763±0.211 | **33.253±0.183** | 33.746±0.361 |
| | Avg | 35.148±1.076 | 34.673±0.855 | 34.129±0.829 | - |
| CR | Lv. 1 | 25.700±0.111 | 25.462±0.212 | 25.184±0.334 | 25.449±0.211 |
| | Lv. 2 | 24.906±0.151 | 24.707±0.228 | 24.465±0.164 | 24.693±0.180 |
| | Lv. 3 | 24.603±0.143 | 24.411±0.282 | **23.710±0.578** | 24.241±0.384 |
| | Avg | 25.070±0.457 | 24.860±0.430 | 24.453±0.602 | - |
| CP | Lv. 1 | 1.975±0.010 | 1.979±0.010 | 1.942±0.012 | 1.965±0.017 |
| | Lv. 2 | 1.930±0.005 | 1.916±0.011 | 1.904±0.008 | 1.917±0.011 |
| | Lv. 3 | 1.916±0.006 | 1.903±0.007 | **1.892±0.008** | 1.904±0.010 |
| | Avg | 1.940±0.025 | 1.933±0.032 | 1.913±0.020 | - |

The combination using Level-1 standardization results for both HS and KSIC Codes exhibited the lowest prediction performance across all EDI states, reporting MAE values of 36.344, 25.700, and 1.975 for the *IN*-, *CR*-, and *CP-states*, respectively. In contrast, the Level-3 combination consistently achieved the best performance, with MAE values of 33.253, 23.710, and 1.892 across all EDI states. Compared to the Level-1 combination, the Level-3 combination achieved performance improvements of 8.50%, 7.74%, and 4.20%. Examining the individual effects of each standardized variable reveals distinct performance trends. KSIC exhibited average MAE values of 35.148, 34.673, and 34.129 for Levels 1–3, respectively, corresponding to a 2.90% improvement at Level-3 relative to Level-1. HS Code standardization yielded MAE values of 35.812, 34.392, and 33.746 across Levels-1–3, resulting in a larger improvement of 5.77% at Level-3 compared to Level-1. These results indicate that increasingly fine-grained standardization of both KSIC and HS Codes provides more informative features for ICDT prediction and that each standardized result independently contributes to performance improvement.

An additional comparative experiment is conducted to verify whether the proposed Gen AI-based standardization is the most effective approach. This experiment adopts the standardization approach of Xie et al. (2025), who proposed a method for standardizing the unstructured information generated by the container terminals. This method embeds standard code descriptions and text data using the same model and the standard code with the highest similarity is the standardization result. This method is applied to the OI and CI using various text embedding methods, including TF-IDF (Salton and Buckley 1988), GloVe (Pennington et al. 2014), and SBERT, a pre-trained text embedding model (Reimers and Gurevych 2019). Additionally, target encoding (Micci-Barreca 2001), a method that can effectively encode high-cardinality categorical variables, is included as the baseline. The experiments are conducted under the same experimental settings as explained in Section 5.1, with a fixed test ratio of 20% for the *IN-state*. The averaged results are summarized in Table 9. Bold and underlined values indicate the best and second-best performing standardization method, respectively, for each model.

**Table 9. Comparison of Text Embedding Techniques**

| | Standardization Methods | | | | | |
|---|---|---|---|---|---|---|
| Model | None | TF-IDF | GloVe | SBERT | Target Encoding | Gen AI |
| XGBoost | 42.113±2.445 | 47.396±5.623 | 43.104±1.110 | 43.531±0.669 | 47.484±6.428 | **37.470±2.818** |
| LightGBM | 38.652±0.105 | 38.566±0.099 | 38.302±0.171 | 37.957±0.162 | 40.501±0.240 | **34.144±0.528** |
| CatBoost | 38.613±0.097 | 38.490±0.181 | 37.807±0.175 | 37.542±0.230 | 39.960±0.457 | **33.253±0.183** |
| SVR | 38.467±0.105 | 38.829±0.218 | 38.743±0.260 | 38.479±0.214 | 38.751±0.272 | **36.409±0.117** |
| MLP | 38.057±0.159 | 40.329±0.137 | 41.410±0.221 | 41.328±0.336 | 40.323±0.451 | **35.212±0.123** |
| FTT | 64.306±0.177 | 63.876±1.543 | **63.875±1.546** | 63.878±1.547 | 63.876±1.547 | 64.302±0.181 |
| NODE | 64.298±0.168 | 63.868±1.546 | **63.864±1.546** | 63.866±1.536 | 63.877±1.548 | 64.296±0.162 |
| TabNet | **62.859±0.967** | 63.875±1.540 | 63.875±1.549 | 63.869±1.553 | 63.875±1.543 | 63.398±1.031 |

For XGBoost, LightGBM, CatBoost, SVR, and MLP, Gen AI-based standardization achieved the best performance. Among the non-Gen AI methods, XGBoost, SVR, and MLP achieved the second-best performance without any standardization, with MAE values of 42.113, 38.467, and 38.057, respectively. LightGBM and CatBoost achieved the second-best performance with SBERT, with MAE values of 37.957 and 37.542, respectively, which were further reduced to 34.144 and 33.253, respectively, corresponding to improvements of 10.05% and 11.42%, respectively, by incorporating Gen AI. For FTT, NODE, and TabNet, the performance differences across standardization methods were negligible. Although pre-trained text embedding models such as SBERT show moderate performance improvements when extracting semantic information from unstructured text, Gen AI consistently achieves the highest performance and extracts the most useful information for ICDT prediction.

Finally, the shapley additive explanations (SHAP) method is employed to analyze the key variables influencing ICDT prediction (Lundberg and Lee 2017). SHAP is an interpretability method based on shapley values from game theory, which quantitatively measures the contribution of each input variable to a model's predictions. It is widely used to interpret complex ML models. The analysis is conducted using CatBoost trained with HS and KSIC codes with Level-3 standardization results, which achieved the best performance as seen in Table 8. Since independent prediction models are trained for each EDI state, variable importance is computed separately for each state. The results are presented in Fig. 4.

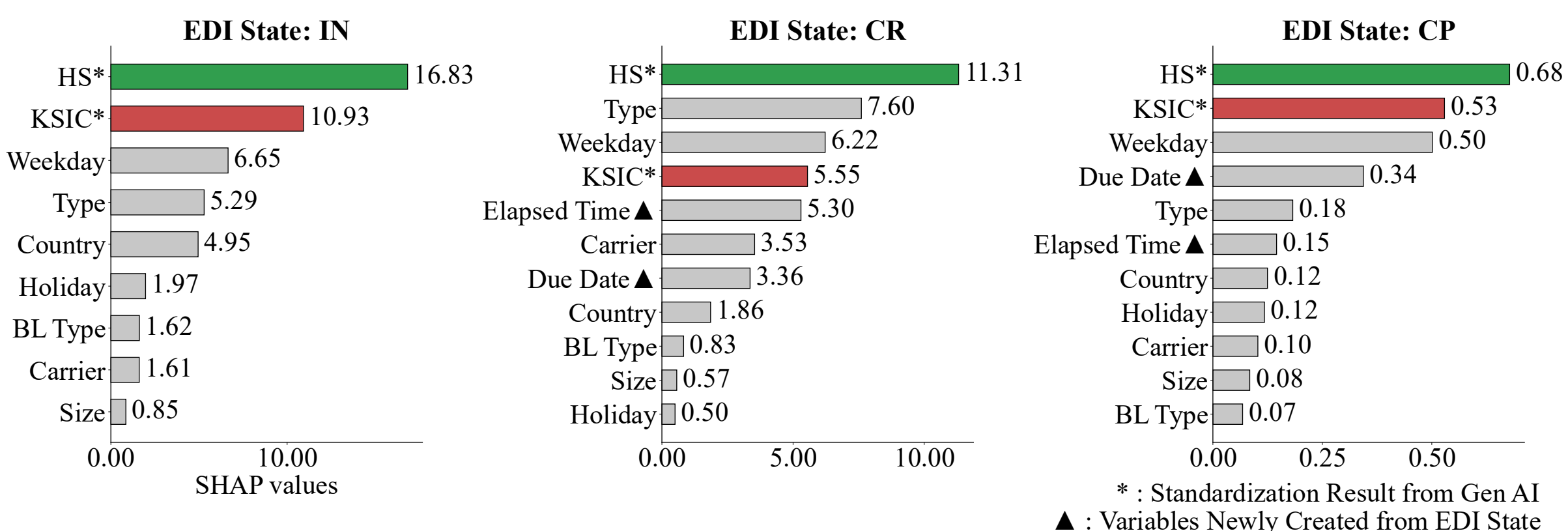

**Figure 4. SHAP Values by EDI State**

In the *IN-state*, *HS* and *KSIC codes* were identified as the most influential variables, followed by *weekday*, *type*, and *country*. In contrast, *holiday*, *bill of lading (BL)*, *type*, *carrier*, and *size* exhibited relatively lower importance. In the *CR-state*, *HS code* remained the most influential variable, while the importance of *KSIC code* slightly decreased. In addition, the $ElapsedTime$ and $DueDate$ variables which are introduced at *CR-state* exhibited contributions to ICDT prediction, indicating that incorporating EDI-based temporal information enhances predictive capability. In the *CP-state*, *HS* and *KSIC codes* again emerged as the most influential variables, while the importance of $DueDate$, which represents the delivery deadline of an import container, increased markedly.

This pattern can be attributed to operational policies requiring the assigned departure truck to complete delivery before the due date.

In addition to the Gen AI-based standardization results, *weekday* and *type* were also identified as variables with high importance. The importance of the variable *weekday* can be attributed to the fact that owners generally conduct their customs clearance and pickup operations on working days. Consequently, containers unloaded on weekends tend to remain in the yard until the next working day, resulting in the ICDT being approximately 1–2 days longer, on average, compared to those unloaded on weekdays. The variable *type*, which represents container categories including reefer and dangerous containers, was identified as the fourth most important feature in the *IN-state* and the second most important feature in the *CR-state*. Because reefer containers typically carry frozen or fresh food products and dangerous containers contain specific chemical materials and other hazardous goods, these variables can implicitly represent the Gen AI-based standardization results. Fig. 5 presents the average ICDT for variables *weekday* and *type*, and provides the empirical evidence for this pattern.

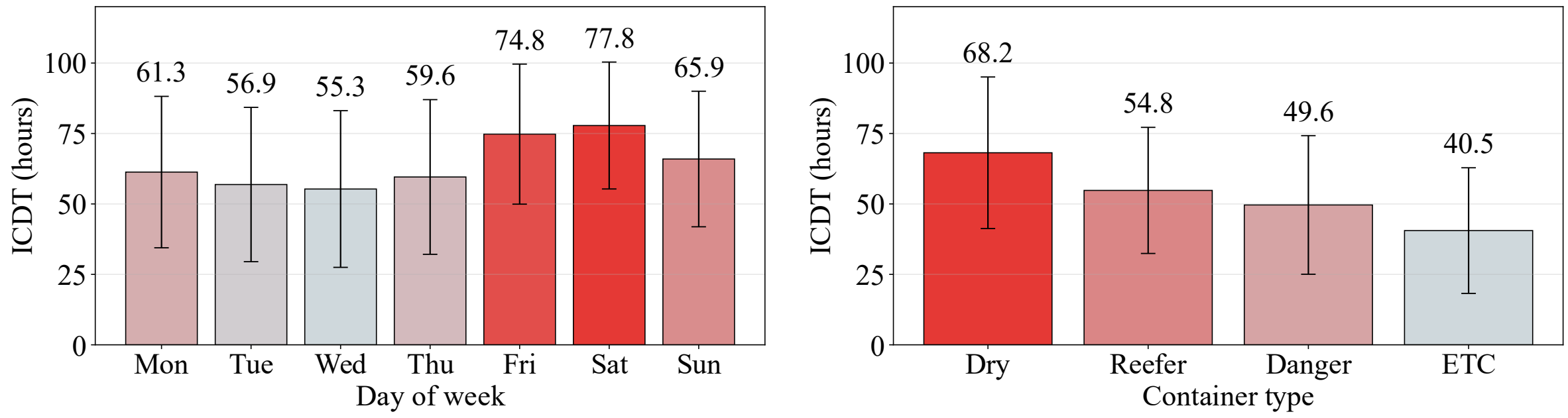

**Figure 5. Average ICDT by variables of weekday and type**

From this perspective, the decrease in the importance of the *KSIC code* from the second most important variable in the *IN-state* to the fourth in the *CR-state* can also be explained. In the *IN-state*, the *KSIC code* is the only variable that captures owner-related information, and the model depends primarily on it. However, in the *CR-state*, $Elapsed\ Time$ and $Due\ Date$ are additionally introduced, both of which are generated only after the cargo owner completes the customs clearance process. The variable *weekday*, defined by the recording time of each EDI state, likewise shifts from denoting the vessel unloading time in the *IN-state* to the customs clearance completion time in the *CR-state*, thereby implicitly representing owner behavior. As a result, owner-related information that was solely captured by the *KSIC code* in the *IN-state* is distributed across four variables, naturally reducing the relative importance of the *KSIC code* alone.

Overall, the Gen AI-based standardization results were consistently identified as highly influential variables across EDI states. Furthermore, variables generated according to EDI state progression were also found to play important roles in ICDT prediction. These findings provide strong evidence that standardizing unstructured data is essential for improving ICDT prediction performance.

## 6. Simulation

This section analyzes whether the two core components of the proposed framework can reduce the number of relocations in the container terminals. Relocation reduction is widely recognized as a critical performance indicator, as it enhances productivity by preventing operational delays through reduced service time and by lowering energy consumption through more efficient equipment utilization (Chhetri et al. 2024; De Armas Jacomino et al. 2021; Maldonado et al. 2019).

To this end, a simulation environment is constructed based on the study by Maldonado et al. (2019), in which the predicted ICDT (p-ICDT) generated by the proposed framework is utilized to determine the stacking positions of containers. This approach is referred to as the *p-ICDT strategy* and is compared with the conventional operational strategy to evaluate whether it reduces the number of relocations. In addition, since the number of relocations may vary depending on yard occupancy, defined as the ratio of currently stacked containers to total yard capacity, multiple occupancy scenarios are constructed by adjusting the available stacking capacity. These scenarios enable a systematic evaluation of the proposed framework under different yard occupancy conditions.

### 6.1 Simulation Setup and Assumptions

To construct a simulation environment that closely reflects real-world operations, the following assumptions are adopted. Conditions (6)–(8) follow the operational practices of the container terminal used in the empirical experiments. In particular, several yard configurations are aligned with actual operational conditions of Busan port, including yard size, maximum stacking height, separate areas for 20 ft and 40 ft containers within the same yard, and a separate yard for reefer containers.

(1) The collected experimental data are sorted by unloading time. 80% of the historical data is used to develop the ICDT prediction models, while the remaining 20% is used for simulation to prevent information leakage. The simulation includes 78,877 import containers that are stacked in the storage yard and subsequently depart from the terminal. The p-ICDT is used to determine their stacking positions, and the simulation terminates once all containers have departed.

(2) During the simulation, ICDT is predicted using CatBoost, which exhibited the best performance in the previous experiments, and these predicted values are used.

(3) Each container is associated with its actual recorded time of unloading, EDI update and departure.

(4) All yards are assumed to have identical dimensions, with 12 rows, 20 bays, and 7 tiers.

(5) A two-week initial loading period is applied, during which containers are stacked in accordance with the conventional operational strategy. The p-ICDT stacking strategy is applied from the third week onward.

(6) Import containers and reefer containers are stacked in separate yards.

(7) Containers of two sizes, 20 ft and 40 ft, are considered. Although both are stored within the same yard, dedicated areas are allocated for each size.

(8) Containers being unloaded must be stacked completely within the current tier before containers can be placed in the next tier.

(9) Containers are randomly assigned to stacking positions in the first tier, while the *p-ICDT strategy* is applied from the second tier onward. This is because the *p-ICDT strategy* requires the presence of containers already stacked in lower tiers to determine appropriate stacking positions.

(10) After unloading, the EDI state of an import container is updated sequentially to IN, CR, CP, and OUT. ICDT is initially predicted at the IN state, then re-predicted at the CR and CP states, and treated as the container's current ICDT at each stage.

(11) The predicted p-ICDT decreases as time elapses, and when it becomes negative, it is fixed at zero.

(12) The number of relocations is defined as the total number of containers stacked above a target container at the time of its departure, with each such container counted as one relocation. All containers stacked above the target container are re-stacked within the same yard according to the same stacking strategy.

**Table 10. Notations for Simulation**

| Notation | Description |
|---|---|
| **Yard Layout** | |
| $y \in \{1, \ldots, Y\}$ | Yard index |
| $r \in \{1, \ldots, R\}$ | Row index |
| $b \in \{1, \ldots, B\}$ | Bay index |
| $t \in \{1, \ldots, T\}$ | Tier index |
| $\hat{t}$ | Tier index of the topmost container currently stacked at position $(y, r, b)$ |
| $P$ | Set of positions across all $(y, r, b, \hat{t})$ |
| **Event & Container** | |
| $i$ | EDI index |
| $k$ | Container index |
| $e_i$ | Type of the $i$-th EDI state, where $e_i \in \{IN, CR, CP, OUT\}$ |
| $c_k$ | Container associated with event $e_i$ |
| $\phi_e(\cdot)$ | ICDT prediction model at EDI state $e_i$ |
| $ICDT_{c_k}$ | p-ICDT of the container $c_k$ |
| **Stacking Decision** | |
| $ICDT_{y,r,b}^{top}$ | p-ICDT for the topmost container at position $(y, r, b)$ |
| $ICDT^*$ | Longest p-ICDT among all topmost containers stacked across $(y, r, b)$ |
| $(y^*, r^*, b^*, t^*)$ | Selected best stacking position |
| $RL$ | Number of relocations |

### 6.2 p-ICDT–Based Stacking Strategy

The *p-ICDT strategy* follows the procedure below. First, when an import container is unloaded, its ICDT is predicted. This predicted value is regarded as the container's p-ICDT until its EDI state is updated. Next, for all operating yard locations (yard, row, bay), the stored p-ICDT of the topmost container is retrieved. The location of the container with the largest p-ICDT (i.e., the container predicted to remain in the yard the longest) is selected, and the new container is stacked on top of it. This strategy minimizes relocations by promoting a stacking order in which containers with imminent departures remain closer to the top. Relocations are counted at the time of an import container's departure from the yard. The overall decision-making process applied to the experimental data is presented in Algorithm 2.

**Algorithm 2. ICDT-Based Yard Stacking Decision Process**

|  |  |  |
|---|---|---|
| 1: | **DEFINITION** | |
| 2: | $RL_{total} = 0$ | # Initialize number of relocations |
| 3: | **GET_ICDT**$(\cdot)$ = retrieve function of ICDT | |
| 4: | **FOR each** $e_i \in \{e_1 \dots e_n\}$ **DO** | # time-ordered sequence of events |
| 5: | $c_k = container\ associated\ with\ e_i$ | |
| 6: | **IF** $e_i = IN$ **THEN** | |
| 7: | $ICDT_{c_k} \leftarrow \phi_{IN}(c_k)$ | # predict ICDT of $c_k$ |
| 8: | $(y^*, r^*, b^*, t^*) = (0,0,0,0)$ | # initialize best position |
| 9: | $ICDT^* = -\infty$ | |
| 10: | $P = \{(y, r, b, \hat{t}) \mid \forall y \in Y, r \in R, b \in B\}$ | |
| 11: | **FOR each** $(y, r, b, \hat{t}) \in P$ **DO** | # search best position |
| 12: | **IF** $\hat{t} + 1 > T$ **THEN** | # check available position |
| 13: | **CONTINUE** | |
| 14: | **ELSE** | |
| 15: | $ICDT^{top}_{y,r,b} =$ **GET_ICDT**$(y, r, b, \hat{t})$ | # retrieve ICDT of container at $(y, r, b, \hat{t})$ |
| 16: | **IF** $ICDT^{top}_{y,r,b} > ICDT^*$ **THEN** | |
| 17: | $ICDT^* \leftarrow ICDT^{top}_{y,r,b}$ | |
| 18: | $(y^*, r^*, b^*, t^*) \leftarrow (y, r, b, \hat{t} + 1)$ | # update best position |
| 19: | **END IF** | |
| 20: | **END IF** | |
| 21: | **END FOR** | |
| 22: | **STACK** $c_k \rightarrow (y^*, r^*, b^*, t^*)$ | # stack $c_k$ at best position |
| 23: | **ELIF** $e_i \in \{CR, CP\}$ **THEN** | |
| 24: | $ICDT_{c_k} \leftarrow \phi_e(c_k)$ | # update ICDT of $c_k$ |
| 25: | **ELIF** $e_i = OUT$ **THEN** | |
| 26: | $(y, r, b, t) = get\ position\ of\ c_k$ | |
| 27: | $\hat{t} = highest\ tier\ at\ (y, r, b)$ | |
| 28: | $RL = \hat{t} - t$ | # calculate relocation |
| 29: | $RL_{total} \leftarrow RL_{total} + RL$ | |
| 30: | **IF** $RL > 0$ ***THEN*** | |
| 31: | $Re - stacking\ containers\ between\ \hat{t}\ and\ t$ | # repeat lines 7–22 |
| 32: | **END IF** | |
| 33: | $remove\ c_k\ from\ the\ yard$ | |
| 34: | **END IF** | |
| 35: | **END FOR** | |

### 6.3 Simulation Results

This subsection considers three experimental settings: (a) an operational baseline, (b) the *p-ICDT strategy*, and (c) the actual-ICDT (a-ICDT) strategy. Setting (a) follows the same simulation setup as described in Section 6.1, but omits the p-ICDT-based stacking decision process. This setting most closely reflects the current operational environment, in which containers are stacked starting from the lowest available tier in the yard. When multiple feasible positions exist at the same tier, one position is randomly selected for stacking. Setting (b) determines the stacking position using p-ICDT values obtained from the proposed framework (Algorithm 2). Setting (c) follows the same decision rule as (b) but replaces p-ICDT with a-ICDT, thereby providing a theoretical upper bound on relocation reduction. When the stacking capacity of the yard is exceeded, containers that cannot be stacked are assumed to be stored in a temporary yard and are excluded from relocation tracking. Table 11 reports the average results obtained over 10 repeated simulations.

**Table 11. Relocation Comparison Across Different Stacking Strategies by Yard Occupancy**

| | Yard Occupancy | | | Number of Relocations | | | Reduction Rate (%) | |
|---|---|---|---|---|---|---|---|---|
| Yard | Init | Avg | Max | (a) Baseline | (b) p-ICDT | (c) a-ICDT | (a)→(b) | (a)→(c) |
| 2 | 66.0% | 72.8% | 100% | 60,672±174 | 59,138±179 | 57,457±171 | 2.53% | 5.30% |
| 3 | 59.4% | 58.3% | 100% | 52,686±104 | 50,567±146 | 47,669±135 | 4.02% | 9.52% |
| 4 | 49.6% | 48.1% | 100% | 44,268±167 | 41,968±82 | 38,257±177 | 5.20% | 13.58% |
| 5 | 42.5% | 39.3% | 86.5% | 34,673±151 | 32,032±80 | 27,895±94 | 7.62% | 19.55% |
| 6 | 37.2% | 32.7% | 72.1% | 26,456±85 | 23,925±104 | 20,107±63 | 9.57% | 24.00% |
| 7 | 33.0% | 28.0% | 61.8% | 20,629±59 | 18,456±95 | 14,697±55 | 10.53% | 28.75% |
| 8 | 29.7% | 24.5% | 54.1% | 16,478±61 | 14,616±62 | 11,234±38 | 11.30% | 31.82% |
| 9 | 27.0% | 21.8% | 48.1% | 13,363±47 | 11,671±54 | 9,085±39 | 12.66% | 32.01% |
| 10 | 24.8% | 19.6% | 43.2% | 10,954±55 | 9,523±39 | 7,428±22 | 13.07% | 32.19% |
| | | | | | | Avg | 8.50% | 21.86% |

The experimental results demonstrate that the *p-ICDT strategy* consistently reduces the number of relocations across all scenarios, achieving an average relocation reduction of 8.50% relative to the baseline. However, the magnitude of this effect differed substantially depending on the yard occupancy. In the Yard 2–4 scenarios, maximum yard occupancy reached 100% at certain time points, whereas the average yard occupancy ranged from 48.1% to 72.8%. This was because yard capacity was temporarily exceeded due to the unloading of large volumes of import containers from ultra-large vessels. Under such high-occupancy conditions, the *p-ICDT strategy* achieved an average relocation reduction of only 3.92%.

As yard occupancy decreased, the relocation reduction effect increased gradually. Specifically, as occupancy decreased from the Yard 5 to Yard 9 scenarios, the relocation reduction rate continuously increased from 7.62% to 12.66%. However, when the average occupancy decreased from 20% to 10%, the improvement plateaued, with the rate changing only from 12.66% to 13.07%. This suggests that below a certain occupancy level, further reductions in relocations are no longer achievable. The target container terminal considered in this study maintains an average yard occupancy of 40%, under which the *p-ICDT strategy* is expected to yield a relocation reduction of 7.62%.

In addition, by comparing the results of a-ICDT and p-ICDT, we analyzed the theoretical upper bound of relocation reduction achievable in the absence of ICDT prediction errors. In the Yard 2 scenario, p-ICDT achieved a relocation reduction of 2.53%, whereas a-ICDT achieved 5.30%, resulting in a difference of only 2.77% between the two methods. Given that a-ICDT is practically unobservable, this result indicates that under high-occupancy conditions, there is limited room for further reduction in relocations even with improved prediction performance. In contrast, the largest gap between p-ICDT and the theoretical upper bound was observed in the Yard 8 scenario, where a-ICDT achieved a 20.52% greater reduction than p-ICDT. This suggests that under relatively low-occupancy conditions, improvements in prediction performance can substantially increase the relocation reduction effect.

These relocation reductions can be further illustrated through the visualization of stacking inversions. A stacking inversion occurs when a lower container has an earlier pickup time than the container stacked above it, necessitating the relocation of the upper container to pick up the lower container. These inversions were identified during simulation by applying the p-ICDT strategy and using a-ICDT as the reference. Fig. 6 visualizes a specific time point during simulation under identical conditions for both strategies, with inversions highlighted in red.

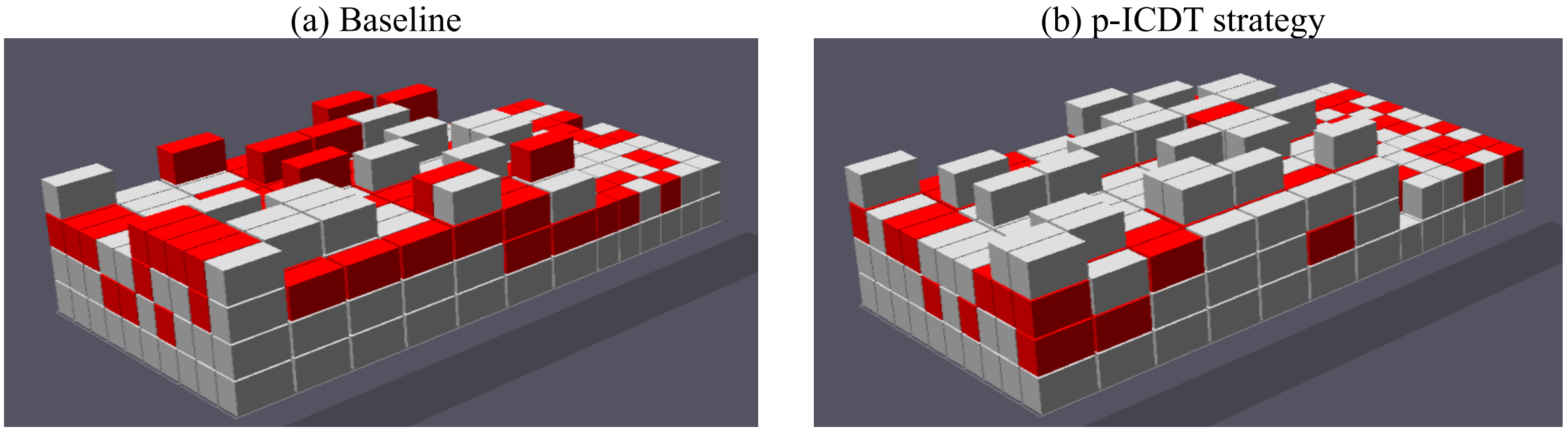

**Figure 6. Visual Comparison of ICDT-based Stacking Strategy and Baseline**

Panel (a) and panel (b) correspond to the experimental settings (a) and (b) defined in Section 6.3, respectively. In this snapshot, panel (b) shows more than a 9.5% reduction in stacking inversions compared to panel (a). This example illustrates how the *p-ICDT strategy* reduces the occurrence of stacking inversions compared to the baseline approach, suggesting its potential to reduce the set of containers requiring relocation.

### 6.4 Effectiveness of Gen AI–Based Standardization

Since this study leverages Gen AI-based standardization results for ICDT prediction, it is important to analyze its impact on productivity improvement. To this end, we apply the *p-ICDT strategy* using CatBoost trained with and without the standardization results and compare the resulting differences in relocation reduction. The results are presented in Table 12.

**Table 12. Impact of Gen AI–Based Standardization on Relocation Reduction**

| Yard | Yard Occupancy | | | Number of Relocations | | Reduction Rate (%) |
|---|---|---|---|---|---|---|
| | Init | Avg | Max | (a) w/o Standardization | (b) w/ Standardization | (a)→(b) |
| 2 | 66.0% | 72.8% | 100% | 59,291±118 | 59,138±179 | 0.26% |
| 3 | 59.4% | 58.3% | 100% | 51,071±141 | 50,567±146 | 0.99% |
| 4 | 49.6% | 48.1% | 100% | 42,522±172 | 41,968±82 | 1.30% |
| 5 | 42.5% | 39.3% | 86.5% | 32,849±84 | 32,032±80 | 2.49% |
| 6 | 37.2% | 32.7% | 72.1% | 24,834±109 | 23,925±104 | 3.66% |
| 7 | 33.0% | 28.0% | 61.8% | 19,221±73 | 18,456±95 | 3.98% |
| 8 | 29.7% | 24.5% | 54.1% | 15,253±58 | 14,616±62 | 4.17% |
| 9 | 27.0% | 21.8% | 48.1% | 12,304±43 | 11,671±54 | 5.15% |
| 10 | 24.8% | 19.6% | 43.2% | 10,121±43 | 9,523±39 | 5.91% |
| | | | | | Avg | 3.10% |

In Table 12, case (a) denotes the application of the *p-ICDT strategy* using a CatBoost trained without the standardization results, whereas case (b) denotes the same strategy applied using a CatBoost trained with the standardization results. Across all scenarios, case (b) consistently achieved fewer relocations than case (a), with an average reduction rate of 3.10%. The magnitude of the effect varied with yard occupancy. Under the conditions where the maximum occupancy reached 100% (Yards 2–4), the average reduction was limited to 0.85%, whereas a value of 4.23% was observed under the remaining conditions. Notably, in Yard 5, whose occupancy level most closely reflects the actual operating conditions of the target container terminal, relocations were reduced by 2.49%.

To observe the daily relocation reduction achieved by applying the proposed framework within the *p-ICDT strategy*, we simulate the settings for Yard 5, which exhibits a yard occupancy level similar to the actual operational environment. The simulation setting is identical to that described in Section 6.1. Two operational situations are observed in detail during the simulation: (1) a week with average yard occupancy and low variability (Fig. 7) and (2) a week in which yard occupancy reaches its maximum during the simulation (Fig. 8). For each situation, we build separate CatBoosts for the cases with and without Gen AI standardization results, apply them to the *p-ICDT strategy*, and compare their relocation performance. Yard occupancy is measured at the end of each day during the simulation and computed as the ratio of stacked containers to the total yard capacity.

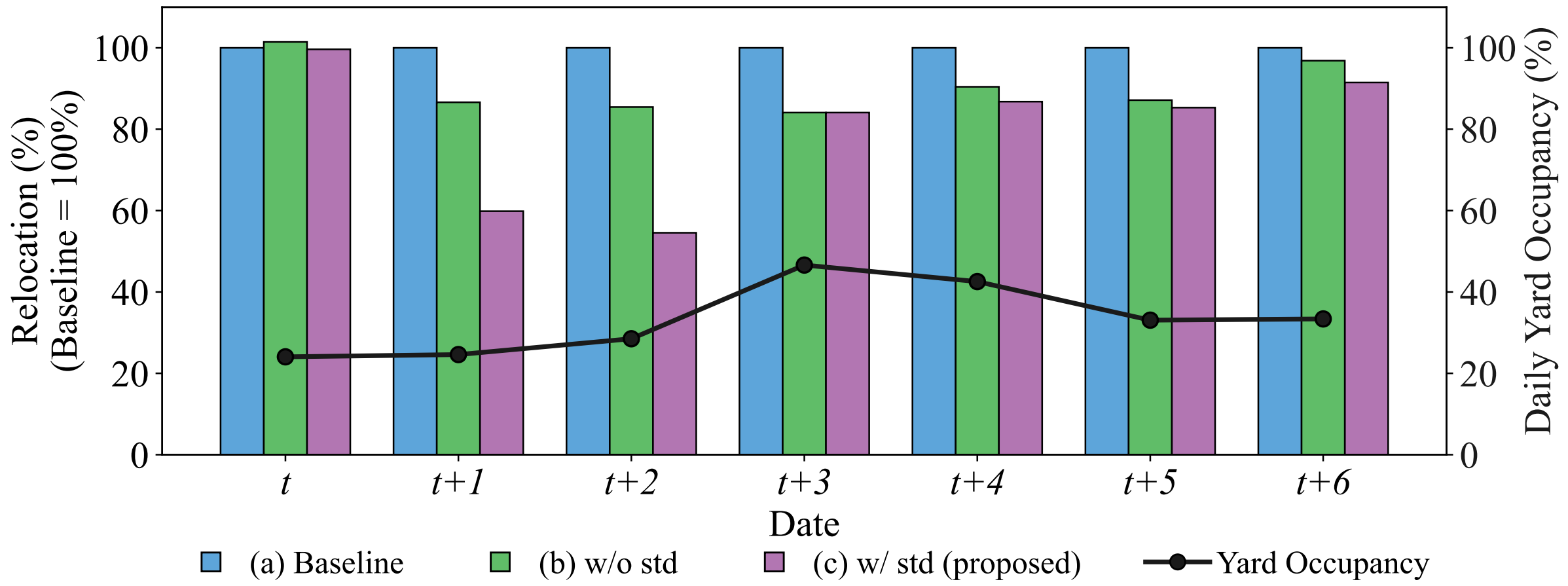

**Figure 7. Relocation Reduction Performance in a Week With Normal Yard Occupancy**

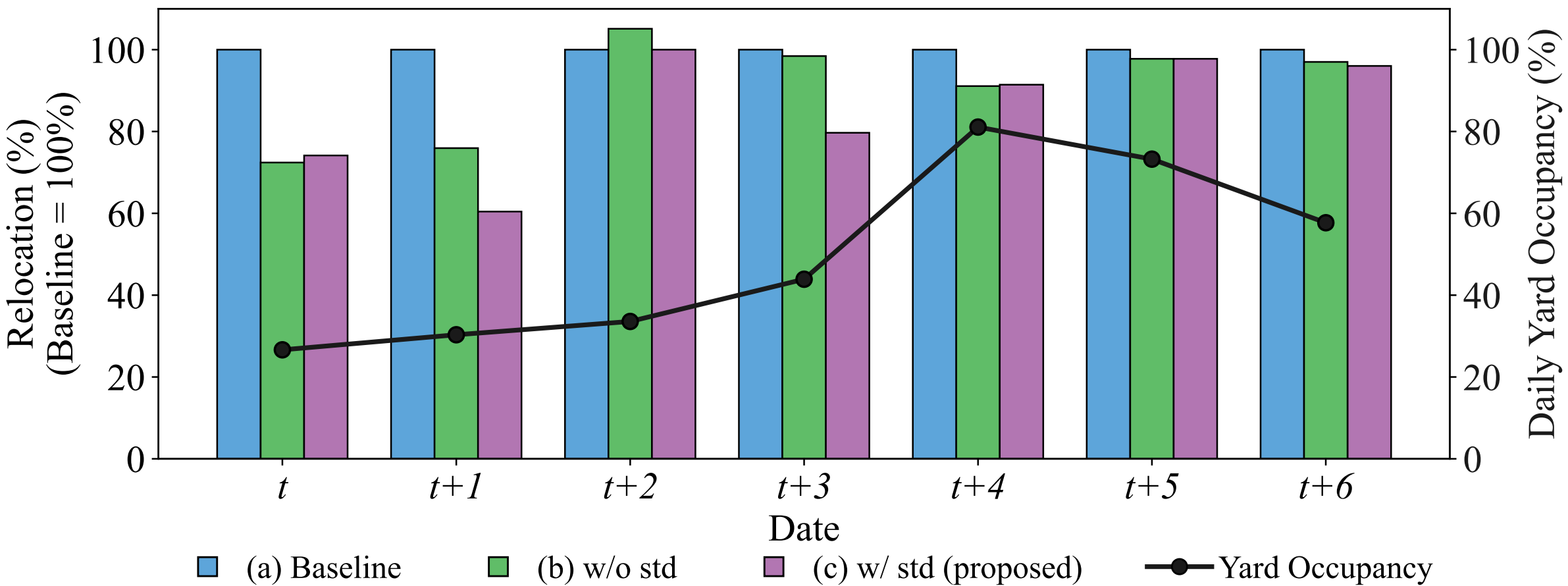

**Figure 8. Relocation Reduction Performance in a Week With the Highest Yard Occupancy**

Fig. 7 presents the results under normal operational conditions, where yard occupancy remains close to its average level. In case (b), which does not use the standardization results, an average relocation reduction of 9.71% was observed relative to the baseline. In contrast, case (c), which uses the standardization results, achieved an average relocation reduction of 19.76%. This demonstrates that, under normal operating conditions, incorporating the Gen AI standardization results as input variables can further reduce relocations even when the same prediction model is used. These findings indicate that adopting Gen AI can make a tangible contribution to improving the operational efficiency of container terminals.

Fig. 8 presents the results for the week with the highest yard occupancy during the simulation period. At times $t + 3$, when yard occupancy was relatively low, case (b) achieved relocation reductions of 1.56% relative to the baseline, whereas case (c) achieved larger reductions of 20.31%, showing a greater effect with the incorporation of standardization results. However, at time $t + 4$, when yard occupancy exceeded 80%, both cases showed nearly identical reductions, indicating that the benefit of using the standardization results becomes negligible at high occupancies. As yard occupancy decreased at times $t + 6$, the pattern re-emerged, with the use of standardization results again leading to additional relocation reductions. These results suggest that Gen AI–based standardization is effective for reducing relocations under typical operational conditions, whereas its effect is limited when yard occupancy is substantially high.

Across both situations, the use of standardization results consistently resulted in fewer relocations than models without standardization. These findings lead to two important implications. First, ICDT predictions generated by the proposed framework can contribute to relocation reduction across diverse operational conditions. Second, to maximize the effect of Gen AI standardization, maintaining yard occupancy at an appropriate level is essential, highlighting a key consideration for applying the *p-ICDT strategy*.

### 6.5 Effectiveness of EDI-Based Re-Prediction

This study re-predicts ICDT whenever the EDI state is updated, based on the assumption that successive updates progressively narrow the plausible range of ICDT and thereby reduce prediction uncertainty. However, if this approach does not contribute to relocation reduction, predicting ICDT only once at the unloading time of an import container may be sufficient. Therefore, we conduct a comparative experiment in which two strategies are examined: (1) predicting ICDT only once at the initial gate-in time and (2) performing EDI-based re-prediction at each state. The results of this comparison are presented in Table 13.

**Table 13. Impact of EDI-Based Re-Prediction on Relocation**

| Yard | (a) Baseline | (b) without EDI | (c) with EDI | Reduction Rate (%) (a)→(b) | Reduction Rate (%) (a)→(c) |
|---|---|---|---|---|---|
| 2 | 60,672±174 | 59,463±195 | 59,138±179 | 1.99% | 2.53% |
| 3 | 52,686±104 | 50,924±160 | 50,567±146 | 3.35% | 4.02% |
| 4 | 44,268±167 | 42,327±133 | 41,968±82 | 4.38% | 5.20% |
| 5 | 34,673±151 | 32,452±142 | 32,032±80 | 6.41% | 7.62% |
| 6 | 26,456±85 | 24,233±96 | 23,925±104 | 8.40% | 9.57% |
| 7 | 20,629±59 | 18,661±85 | 18,456±95 | 9.54% | 10.53% |
| 8 | 16,478±61 | 14,719±81 | 14,616±62 | 10.68% | 11.30% |
| 9 | 13,363±47 | 11,771±57 | 11,671±54 | 11.91% | 12.66% |
| 10 | 10,954±55 | 9,590±58 | 9,523±39 | 12.45% | 13.07% |
| | | | Avg | 7.68% | 8.50% |

Under the condition without EDI, an average relocation reduction of 7.68% was observed relative to the baseline, whereas the condition with EDI achieved 8.50%, indicating that the EDI-based re-prediction provides an additional average reduction of 0.82%. In particular, under the Yard 5 setting, which is similar to the current operational environment, this approach yielded an additional reduction of 1.21%. This improvement can be attributed to improved prediction accuracy associated with EDI state transitions. As shown in Table 7 in Section 5.2, MAE decreased from 33.253 to 23.710 when transitioning from the *IN-state* to the *CR-state*, indicating that the availability of customs clearance completion information reduces uncertainty in ICDT prediction. Because the re-predicted ICDT reflects the imminence of container retrieval more accurately, the *p-ICDT strategy* can make more precise decisions. Moreover, if additional EDI states are identified through collaboration with domain experts in future work, further improvements in ICDT prediction performance and relocation reduction are expected.

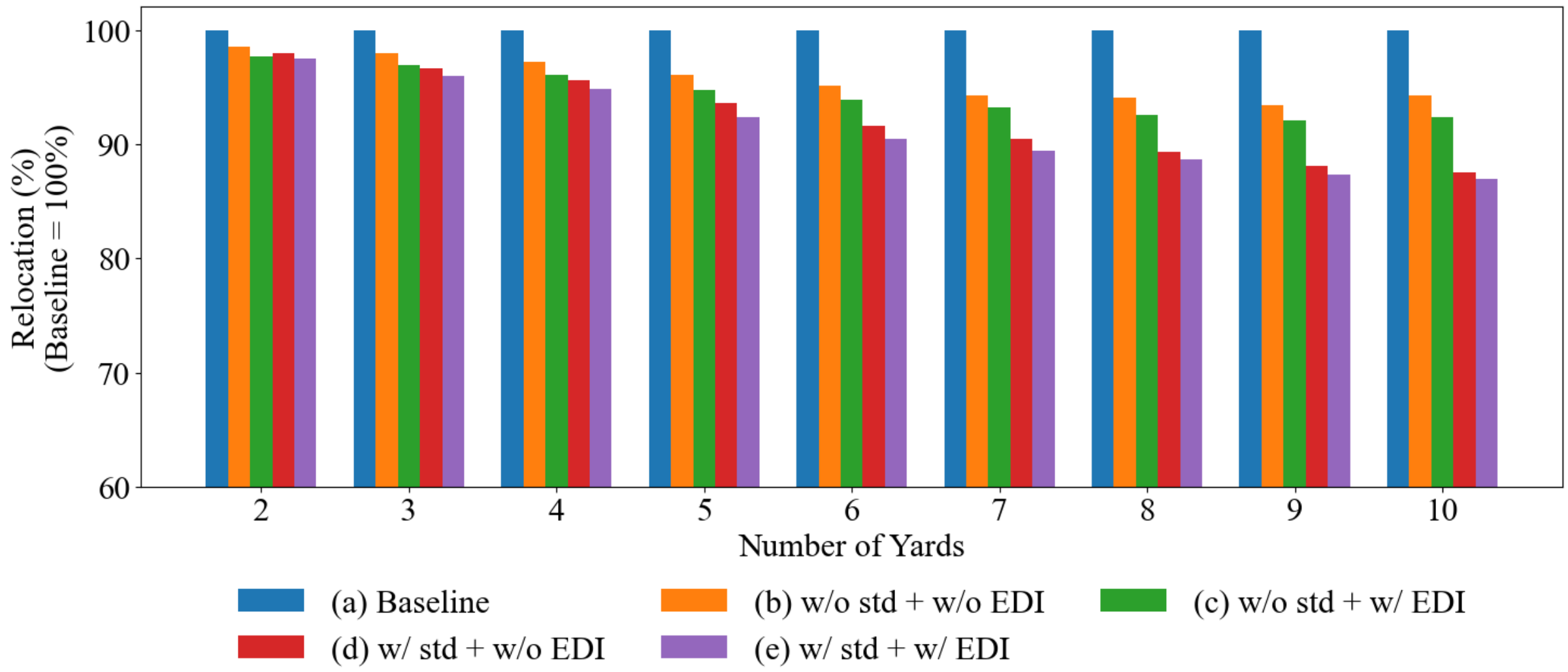

**Figure 9. Relocation Reduction Across Different Combinations of Core Components**

Finally, we comprehensively compare the relocation reduction effects associated with the two core components of the proposed framework. Fig. 9 shows the relocation reduction ratios under different component configurations, with the number of relocations in the operational baseline (a) normalized to 100%. In configuration (b), ICDT is predicted without the use of standardization results and without EDI-based re-prediction. Configuration (c) applies only EDI-based re-prediction, configuration (d) utilizes only the standardization results, and configuration (e) represents the full proposed framework, incorporating both components. The results show a consistent pattern of progressively decreasing relocations in the order of (b)→(c)→(d)→(e). Compared with the baseline, (b) achieved an average relocation reduction of 4.35% across all yard settings. Applying only EDI-based re-prediction in (c) achieved 5.61%, while using only the standardization results in (d) achieved 7.68%. When both components were applied in (e), an average relocation reduction of 8.50% was achieved compared with the baseline. Under the Yard 5 setting, which is the most similar to the operational environment, relocation reductions relative to the baseline increased monotonically from 3.97% in (b), to 5.26% in (c), 6.41% in (d), and 7.62% in (e). These results quantitatively demonstrate the effects of the two core components and confirm that the proposed framework can make a concrete and measurable contribution to productivity improvement in container terminals.

## 7. Discussion

### 7.1 Applicability of the Collaborative Framework

The proposed framework is designed with practical applicability to container terminals and offers four advantages. 1) It can be easily integrated with existing terminal operation system databases. Container terminals typically receive advance information on containers scheduled for unloading, enabling standardization to be performed in advance using Gen AI. This design resolves potential bottlenecks that may arise when deploying Gen AI in real-time operational settings and, when combined with the fast inference time of the ML model, allows the framework to operate without undue burden in practice. Moreover, because the framework follows an end-to-end structure that operates without raw data processing or additional preprocessing, dedicated maintenance personnel are not required. This feature enables conservative practitioners in container terminals to adopt state-of-the-art technology without changing the existing operational environment. 2) The framework ensures interpretability of both the standardization and prediction results. Since both the standardization process (see Tables 4–6) and the ML model are explicitly interpretable, the framework provides transparent explanations of variable contributions. This high level of interpretability enables port operators to establish various policies based on the interpreted results. 3) The framework supports multilingual operation. Port logistics is an international industry, and OI/CI are recorded and managed in diverse languages. Gen AI enables these heterogeneous textual inputs to be consistently mapped to unified standard codes, eliminating linguistic barriers. 4) The caching mechanism of the STD Bank continuously reduces operational costs as standardization results accumulate over time (see Fig. 10).

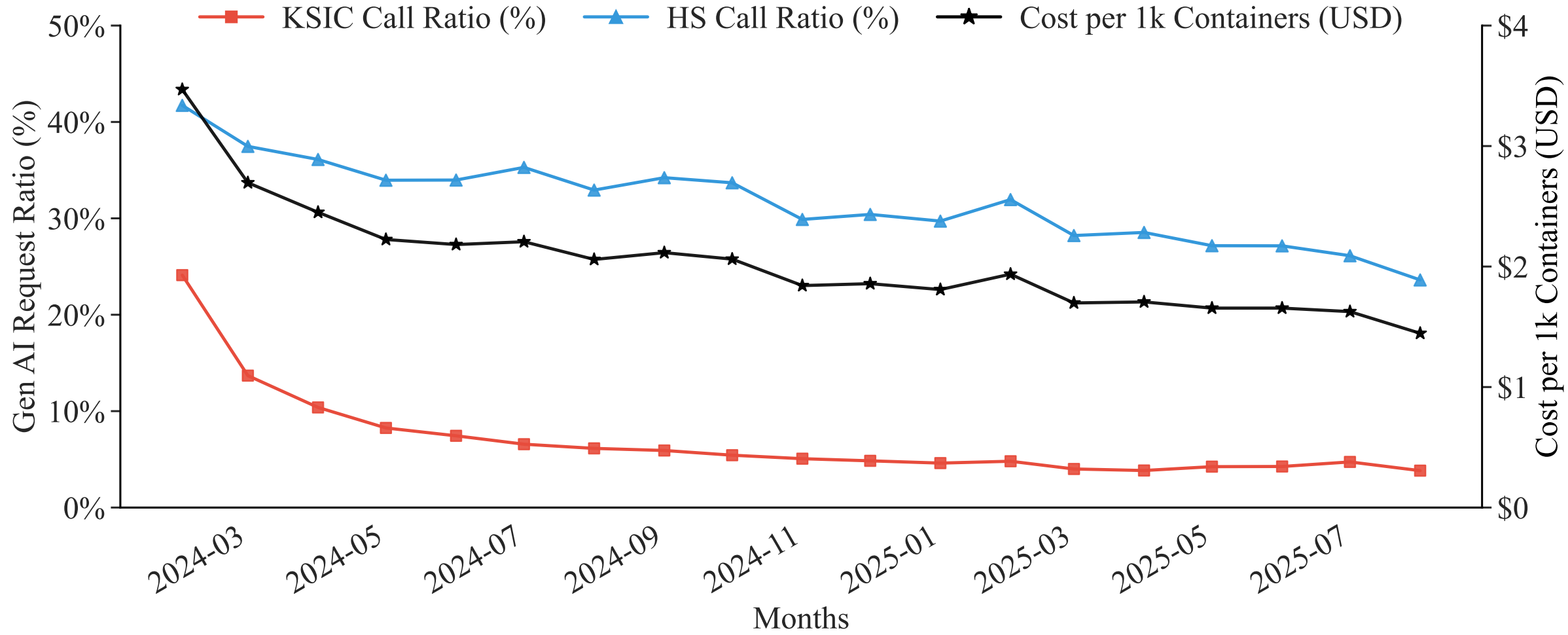

**Figure 10. Progressive Reduction of Gen AI Request Ratio and Cost**

Fig. 10 illustrates the Gen AI usage ratio and the associated cost per 1,000 containers, where the cost is based on Gemini 2.5 Flash, which was adopted in this study. In the first month after introducing the proposed framework, Gen AI was used for 26% of OI and 40% of CI among all unloaded import containers, resulting in an approximate cost of 3.8 USD per 1,000 containers. After one year, the Gen AI usage ratio decreased to 4.82% for OI and 29.92% for CI, reducing the cost to 1.92 USD per 1,000 containers. For CI, the Gen AI call ratio decreased more gradually than that for OI. This is attributable to the higher variability in CI expressions compared with OI. This trend is particularly important from an economic perspective. While relocation-related operational costs accumulate approximately linearly over time as yard operations continue, the cost associated with Gen AI usage decreases due to the caching mechanism of the STD Bank. Consequently, a break-even point can emerge at which the cumulative savings achieved through relocation reduction offset the cost incurred by adopting Gen AI. In summary, the proposed framework offers high practicality due to its deployability, interpretability, multilingual capability, and cost-effectiveness. These characteristics indicate that the framework can serve as a useful reference for integrating Gen AI into real-world port logistics operations.

### 7.2 Practical Considerations

This study employed an EDI-tracking and re-prediction strategy to reduce uncertainty in ICDT prediction. However, because EDI systems can differ across countries and ports, applying the proposed framework in practice requires identifying EDI events that are suitable for the target terminal in collaboration with domain experts. For example, when a Terminal Appointment System (TAS) is available, the framework can initially predict ICDT at the unloading stage, then refine the prediction once the pickup truck's expected arrival time is confirmed through TAS. Incorporating such port-specific characteristics plays a key role in the successful application of the proposed framework in practice.

In addition, because ICDT inherently involves high uncertainty, securing reliable training data is essential. This requires establishing systematic operational standards and close collaboration among stakeholders. High-quality data can be obtained under sustainable operational regulations, under which variables maintain meaningful relationships with ICDT. In contrast, in environments where such policies are absent or poorly designed, it may be difficult to collect data with predictable patterns. In some ports, ICDT has been reported to extend up to 90 days due to excessively low storage fees. Under such conditions, ICDT is determined by stakeholders' arbitrary decisions rather than by relationships among variables, thereby constraining the development of reliable prediction models. Therefore, to ensure the effectiveness of ICDT prediction, an operational system that guarantees the predictability of ICDT should be established before adopting prediction technologies.

Finally, operational data distributions in container terminals can shift over time owing to various reasons. These shifts can gradually degrade prediction performance, particularly because the model should predict future ICDT based solely on past observations. This can pose a significant challenge to the reliability of the deployed prediction model. To maintain reliable prediction in a real operational environment, the model must be periodically updated by incorporating newly accumulated data through strategies such as online learning. This strategy can sustain prediction performance under shifting operational data distributions.

### 7.3 Security and Privacy Considerations

Container terminals serve as intermediaries in supply chains and handle large volumes of sensitive information. Because CI contains detailed information on imported items and OI identifies the entity associated with those items, processing such data with Gen AI may raise security concerns among terminal operators and their customers. To address this issue, two approaches can be considered. First, Gen AI infrastructure can be established at the port authority level. Compared with individual terminal operators directly using commercial Gen AI services, infrastructure built and managed by a public institution provides greater neutrality and trust. Second, Gen AI can be deployed on-premise within individual container terminals. In this case, data remain within the terminal and are processed internally, thereby eliminating the risk of external information leakage. These approaches can enable individual terminal operators to utilize advanced technologies while addressing security concerns, potentially facilitating broader adoption of Gen AI technologies across the port logistics industry.

### 7.4 Limitations

A limitation of this study is that Gen AI does not guarantee consistent outputs for the same input. This is confirmed by the consistency test results. Table 14 reports an experiment in which the top 100 most frequent entries from CI and OI are extracted, and the same inputs are repeatedly provided to examine whether consistent standardization results are obtained. The consistency rate is computed using Eq. (21).

$$Consistency\ Rate = \left(1 - \frac{N_{unique} - 1}{N_{total} - 1}\right) \times 100 \tag{21}$$

**Table 14. Consistency Rate of CI and OI Standardization**

| Information | Standardization Level | | |
|---|---|---|---|
| | Level-1 | Level-2 | Level-3 |
| CI | 98.44% | 96.63% | 90.31% |
| OI | 96.15% | 91.25% | 84.05% |

At Level-1, a consistency rate of 98.44% was observed, meaning that approximately 98 out of the 100 CI entries returned identical standardization results across 10 repeated requests. At Level-2 and Level-3, the consistency rates were 96.63% and 90.31%, respectively. For OI, a high consistency rate of 96.15% was observed at Level-1, whereas the rates decreased to 91.25% and 84.05% at Level-2 and Level-3, respectively.

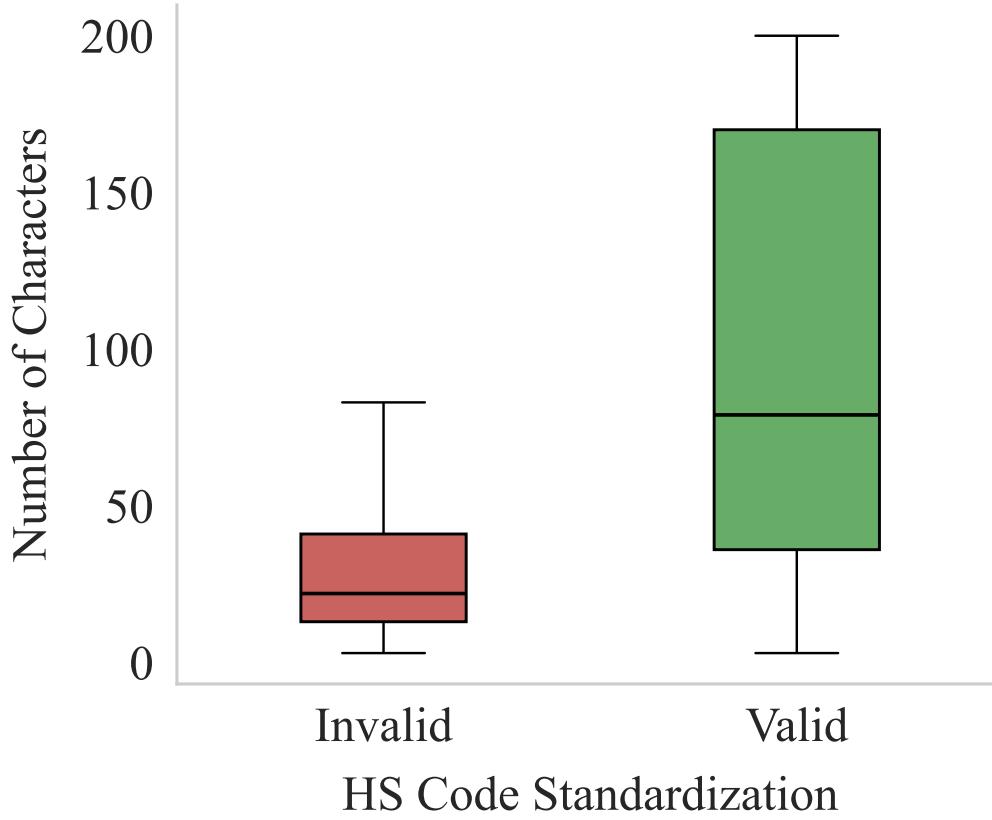

**Figure 11. Distribution of CI Text Length**

**Table 15. Examples of Gen AI Reasoning for Invalid Classification**

| CI | Reason* |
|---|---|
| TUBE | The cargo description "TUBE" is a very broad term that lacks specific attributes such as material, usage, or diameter. Based on this information alone, it is not possible to identify the corresponding HS Code at the 2-, 4-, or 6-digit level. Therefore, additional information is required for accurate HS Code classification. |
| STERILE | The cargo description "STERILE" only indicates the condition or characteristic of the item and does not provide specific product names or material information. Therefore, it is not possible to classify it under any HS Code. Additional cargo details that reveal the essential nature of the product are required for accurate HS Code classification. |
| GARMENTS | The cargo description "GARMENTS" alone lacks the detailed information necessary for HS Code classification at the 6-digit level, such as material composition, product type (e.g., shirts, trousers), and whether the item is knitted or woven. Therefore, precise HS Code classification based on the essential characteristics of the goods is not possible. |

***: Generated from Gen AI**

Such reductions in consistency rate tend to occur when the provided information is insufficient. Fig. 11 illustrates this pattern. When Gen AI judged standardization to be feasible during the validation check, the input text had an average length of 85 characters. Meanwhile, when it was judged as invalid, the average length was approximately 20 characters, which is more than four times shorter than those in the valid cases. This finding indicates that shorter input texts often fail to provide sufficient information for Gen AI to determine the HS code, leading to limitations in standardization. This observation is further supported by the Gen AI reasoning results in Table 15. For example, for extremely short CI entries such as "TUBE," "STERILE," and "GARMENTS," Gen AI provided reasons such as "lacks specific attributes," indicating that inputs were too ambiguous to identify the essential characteristics of the product. Therefore, improving standardization consistency under limited-information conditions remains a key limitation of this study and represents an important direction for future research. Specifically, prompt engineering techniques can be refined to guide the model toward more precise standardization outputs. Retrieval-augmented generation can be adopted to enable Gen AI to directly reference official classification manuals such as the HS code explanatory notes during standardization. Furthermore, employing more capable alternative models with stronger reasoning abilities could enhance the consistency of the standardization results.

## 8. Conclusion and Future Research

The proposed framework improves ICDT prediction performance by combining Gen AI-based standardization and EDI-based re-prediction, achieving up to a 13.88% improvement in ICDT prediction. In addition, when the predicted ICDT values were incorporated into a container stacking strategy, it achieved an average relocation reduction of 8.50% and a maximum reduction of 13.07% relative to the operational baseline. Notably, even when all other variables and the same prediction model were used, incorporating the Gen AI-based standardization results led to an additional average relocation reduction of 3.10%. These results have important implications in that they identify concrete points at which Gen AI can be applied in container terminal operations and empirically demonstrate that Gen AI can contribute to productivity improvement.

This study has important implications not only for productivity improvement but also for environmentally sustainable operations. Modern port operations are increasingly required to simultaneously achieve productivity and environmental performance; however, these objectives are often in a trade-off relationship. Container terminals tend to maximize equipment utilization to prevent delays in vessel operations, which can improve productivity but may conflict with environmental sustainability. A practical way to mitigate this dilemma is to improve equipment-use efficiency for the same workload. By reducing unnecessary relocations through ICDT-based stacking strategies, the proposed framework can reduce unnecessary yard crane operations for a given container workload. This suggests a viable pathway for simultaneously improving productivity and environmental performance, thereby contributing to sustainable port operations.

Future work can extend this study by addressing the limitations discussed in Section 7.4. First, further studies are needed to improve the consistency of standardization. Specifically, this may involve advancing prompt-engineering techniques and introducing multi-agent approaches to validate standardization results. In addition, Retrieval-Augmented Generation can be adopted to enable standardization with direct access to the target classification systems. Second, developing resource-efficient Small Language Models with performance comparable to commercial Gen AI represents an important research direction for container terminal applications. Such models would enhance practical deployability while also addressing security concerns. Third, Gen AI has the potential to extract additional useful information from unstructured text beyond standard code systems, such as distinguishing between full container load (FCL) and less than container load (LCL), identifying seasonal product characteristics from CI, or investigating import patterns of specific commodity types. Exploring such features represents a promising direction for extracting richer representations from unstructured data and further enhancing ICDT prediction performance. Finally, because this study was conducted at a single container terminal at Busan port, future work should validate the generalizability of the proposed framework across diverse port environments. Through these follow-up studies, Gen AI-based approaches are expected to become practical solutions that improve both productivity and sustainability across the port logistics domain.

**Appendix 1. Prompt of KSIC code Standardization**

**[Instruction]**

You are a Port EDI Data Standardization Assistant and an expert in corporate industrial classification. Your role is to perform the following tasks based on the provided information:

1. Utilize owner information included in imported container records, along with relevant web search results.
2. Generate the output strictly according to the provided JSON schema.

**[Task Description]**

**1) KSIC Classification (Korean Standard Industrial Classification)**

- Based on the inferred information, classify the company according to the KSIC system.
- The KSIC is Korea's national standard for industrial classification, derived from the UN's International Standard Industrial Classification (ISIC) and adapted to reflect the Korean industrial context.

**2) Validation Check**

- assign the validation check field to exactly one of the following values: "type1", "type2", or "type3".

type1: the provided information is valid, and you are able to standardize it with sufficient confidence.

type2: the provided information is partially valid, but it is insufficient or ambiguous, resulting in low confidence in standardization.

type3: the provided information is invalid, or cannot be standardized under any condition.

**3) Standardization**

- High Level (Section): Represents the broadest industrial category, denoted by a single uppercase letter (A–U).
- Middle Level (Division): Two-digit numeric code that subdivides the Section.
- Low Level (Group): Three-digit numeric code that further subdivides the Division.
- Classify the company size as one of the following categories: SME, Mid, Large, or Unknown (use Unknown when information is unavailable).

**Output Requirements**

- The output must be a single JSON object strictly following the provided schema.
- No additional fields are allowed.
- All key names and structures must match the schema exactly.

**JSON Scheme**

```
{
  owner: Input name of owner,
  size: Company size classification: one of ["SME", "Mid", "Large", "Unknown"] (use "Unknown" when information is unavailable),
  section1: High-level KSIC code (A–U, single uppercase letter),
  division2: Middle-level KSIC code (2 digits),
  group3: Low-level KSIC code (3 digits),
  validation_check: type1: valid information / type2: insufficient information / ksic-type3: invalid information,
  reason: Explanation of the KSIC standardization result
}
```

**[Request]**

Convert the following input information into JSON format according to the provided guidelines.

Owner Information: $OI_{raw}$

## Appendix 2. Prompt of HS code Standardization

**[Instruction]**

You are a Port EDI Data Standardization Assistant and an expert in product classification. Your role is to perform the following tasks based on the provided information:

1. Standardize each cargo information according to the HS Code (Harmonized System) at the 2-, 4-, and 6-digit levels, following the official HS system.
2. Produce the output strictly in accordance with the provided JSON schema.

**[Task Description]**

**1) HS Code Classification**

- Determine the HS Code (2-, 4-, and 6-digit levels) based on the cargo's essential nature (material/composition), primary function or use, degree of processing or form, and whether it is a finished or part item.

**2) Validation Check**

- assign the validation check field to exactly one of the following values: "type1", "type2", or "type3".

type1: the provided information is valid, and you are able to standardize it with sufficient confidence.

type2: the provided information is partially valid, but it is insufficient or ambiguous, resulting in low confidence in standardization.

type3: the provided information is invalid, or cannot be standardized under any condition.

**3) Standardization**

- High Level (2-digit Chapter): Represents the broadest category, defined by the first two digits of the HS code.
- Middle Level (4-digit Heading): A four-digit numeric code that subdivides the Chapter, and must begin with the same two digits as the High Level.
- Low Level (6-digit Subheading): A six-digit numeric code that further subdivides the Heading, and must begin with the same four digits as the Middle Level.

**Output Requirements**

- The output must be a single JSON object strictly following the provided schema.
- No additional fields are allowed.
- All key names and structures must match the schema exactly.

**JSON Scheme**

{

cargo: The original cargo information exactly as entered (must not be modified, added, deleted, or reformatted in any way, including changes in case or spacing),

hscode2: Two-digit HS Chapter code; enter null only if classification is not possible,

hscode4: Four-digit HS Heading code; enter null only if classification is not possible,

hscode6: Six-digit HS Subheading code; enter null only if classification is not possible,

evidence_tokens: Key terms or tokens that influenced classification (e.g., "cotton 100%", "air conditioner parts"); enter an empty array if classification was not possible,

validation_check: type1: valid information / type2: insufficient information / type3: invalid information,

reason: Explanation of the HS standardization result

}

**[Request]**

Standardize the following input information into JSON format according to the provided guidelines.

Container Information: $CI_{raw}$

**Appendix 3. Abbreviations**

| Abbreviation | Full Name | Abbreviation | Full Name |
|---|---|---|---|
| ICDT | Import Container Dwell Time | TF-IDF | Term Frequency–Inverse Document Frequency |
| p-ICDT | Predicted Import Container Dwell Time | BERT | Bidirectional Encoder Representations from Transformers |
| a-ICDT | Actual Import Container Dwell Time | SBERT | Sentence-BERT |
| CI | Cargo Information | GloVe | Global Vectors for Word Representation |
| OI | Owner Information | XGBoost | Extreme Gradient Boosting |
| EDI | Electronic Data Interchange | LightGBM | Light Gradient Boosting Machine |
| HS Code | Harmonized System Code | CatBoost | Categorical Boosting |
| NAICS | North American Industry Classification System | SVR | Support Vector Regression |
| KSIC | Korean Standard Industrial Classification | MLP | Multi-Layer Perceptron |
| BL | Bill of Lading | FTT | FT-Transformer |
| DO | Delivery Order | NODE | Neural Oblivious Decision Ensembles |
| SME | Small and Medium-sized Enterprise | TabNet | Attentive Interpretable Tabular Learning |
| TAS | Terminal Appointment System | MAE | Mean Absolute Error |
| STD Bank | Standardization Bank | Gen AI | Generative Artificial Intelligence |
| ML | Machine Learning | LLM | Large Language Model |
| DL | Deep Learning | NLP | Natural Language Processing |
| SHAP | Shapley Additive Explanations | | |

**Appendix 4. Configurations of Gen AI (Gemini 2.5 Flash)**

| Parameter | Value | Description |
|---|---|---|
| temperature | 1 | Value that controls the degree of randomness in token selection. |
| top_p | 0.95 | Tokens are selected from the most to least probable until the sum of their probabilities equals this value. |
| top_k | 64 | For each token selection step, the top_k tokens with the highest probabilities are sampled. Tokens are then further filtered based on top_p, with the final token selected using temperature sampling. |
| max_output_tokens | - | Maximum number of tokens that can be generated in the response. |
| candidate_count | 1 | Number of response variations to return. |